\documentclass[12pt,onecolumn,oneside]{IEEEtran}
 
\usepackage{amsmath}
\usepackage{amssymb}
\usepackage{dsfont}
\usepackage{epsfig}
\usepackage{cite}

\newcommand{\comment}[1]{}

\newtheorem{thm}{Theorem}

\newtheorem{cor}[thm]{Corollary}

\newcommand{\reltxt}[2]{\stackrel{\text{\tiny #1}}{#2}}
\newcommand{\relas}[1]{\reltxt{a.s.}{#1}}

\DeclareMathOperator*{\arginf}{arg \mspace{3mu} inf}

\DeclareMathOperator*{\Prob}{Prob}
\DeclareMathOperator*{\ave}{ave}

\newcommand{\ind}{\mathds{1}}
\newcommand{\matsize}[2]{{#1}\mspace{-3mu}\times\mspace{-3mu}{#2}}

\newcommand{\Dmin}{D_{\min}}
\newcommand{\Dave}{D_{\ave}}

\renewcommand{\theenumi}{\roman{enumi}} 

\def\be{\begin{eqnarray}}
\def\ee{\end{eqnarray}}
\def\ben{\begin{eqnarray*}}
\def\een{\end{eqnarray*}}

\renewcommand{\relas}[1]{\reltxt{w.p.1}{#1}}

\newcommand{\Dc}{D_{\text{c}}}

\begin{document}
\renewcommand{\textfraction}{0}
 
\title{Estimation of the Rate-Distortion Function}

\author{Matthew T.~Harrison,~\IEEEmembership{Member,~IEEE} and
Ioannis Kontoyiannis,~\IEEEmembership{Senior Member,~IEEE}
\thanks{A shorter version of this paper is to appear in {\it IEEE Transactions on Information Theory}.}
\thanks{MH was supported in part by a National Defense Science and Engineering Graduate Fellowship.   IK was supported in part by a Sloan Research Fellowship from the Sloan Foundation, NSF grant \#0073378-CCR and USDA-IFAFS grant \#00-52100-9615.  The material in this paper is preceded by two technical reports \cite{Har:Epi:2003,Har:Convergence:2003}.  Preliminary results were presented at \cite{HarKon:Maximum:2002} and \cite{HarKon:Estimating:2006}.}
\thanks{MH is at the Department of Statistics, Carnegie Mellon University, Pittsburgh, PA 15213 (email: mtharris@cmu.edu).}
\thanks{IK is at the Department of Informatics, Athens University of Econ \& Business, Patission 76, Athens 10434, Greece (email: yiannis@aueb.gr).}}

\maketitle

\begin{abstract}
Motivated by questions in lossy data compression 
and by theoretical considerations, we
examine the problem of estimating the rate-distortion 
function of an unknown (not necessarily discrete-valued)
source from empirical data. Our focus 
is the behavior of the so-called ``plug-in'' estimator, 
which is simply the rate-distortion function of the empirical 
distribution of the observed data. Sufficient conditions 
are given for its consistency, and examples are provided
to demonstrate that in certain cases it fails to converge 
to the true rate-distortion function.
The analysis of its performance is complicated by the fact
that the rate-distortion function is not continuous in the source
distribution; the underlying mathematical problem is closely related 
to the classical problem of establishing the consistency 
of maximum likelihood estimators.
General consistency results are given for the plug-in estimator
applied to a broad class of sources, including all
stationary and ergodic ones.
A more general 
class of estimation problems is also considered, arising in the 
context of lossy data compression when the allowed class 
of coding distributions is restricted; analogous results are 
developed for the plug-in estimator in that case.
Finally, consistency theorems are formulated for 
modified (e.g., penalized) versions of the plug-in,
and for estimating the optimal reproduction distribution.
\end{abstract}

% instances where 
% the rate-distortion function is defined over a restricted class of coding 
% distributions, and estimation of an optimal output distribution.

% The problem of estimating the entropy of a discrete, memoryless source from 
% data has received a lot of attention.  In this work we consider the natural 
% generalization of estimating the rate-distortion function.  Our motivation 
% comes from questions in lossy data compression and from cases where the data 
% under consideration do not take values in a discrete alphabet.  
% Our primary 
% focus is the asymptotic behavior of the plug-in estimator that uses the 
% rate-distortion function of the empirical distribution of the data as an 
% estimate.  In most (but not all) cases this estimator is strongly consistent, 
% that is, it converges to the true value with probability one as the amount 
% of data increases.  


\IEEEpeerreviewmaketitle

%%%%%%%%%%%%%%%%%%%%%%%%%%%%%%%%%%%%%%%%%%%%%%%%%
%%%%%%%%%%%%%%%%%%%%%%%%%%%%%%%%%%%%%%%%%%%%%%%%%
%%
% %  INTRODUCTION
%%
%%%%%%%%%%%%%%%%%%%%%%%%%%%%%%%%%%%%%%%%%%%%%%%%%
%%%%%%%%%%%%%%%%%%%%%%%%%%%%%%%%%%%%%%%%%%%%%%%%%

\section{Introduction}

Suppose a data string $x_1^n := (x_1,x_2,\dotsc,x_n)$ is generated 
by a stationary memoryless source $(X_n\,;\,n\geq 1)$
with unknown marginal distribution $P$ on a discrete
alphabet $A$. In many theoretical and practical 
problems arising in a wide variety of scientific
contexts, it is desirable -- and often important --
to obtain accurate estimates of 
the entropy $H(P)$ of the source, based on the 
observed data $x_1^n$; see, for example, 
\cite{schu-grass:96}\cite{kasw}\cite{LW:99}\cite{levene-loizou:00}%
\cite{paninski:03}\cite{nemenman:04}\cite{C-K-Verdu:04}
and the references therein.
% Much is known about this problem.
Perhaps the simplest method 
is via the so-called {\bf plug-in estimator}, 
where the entropy of $P$ 
is estimated by 
$H(P_{x_1^n})$,
namely, the entropy 
of the empirical distribution 
$P_{x_1^n}$
of $x_1^n$.
The plug-in estimator satisfies the basic statistical 
requirement of consistency, that is, 
$H(P_{X_1^n}) \to H(P)$ in probability as $n\to\infty$.  
In fact, it is strongly consistent; the convergence 
holds with probability one%, and moreover it holds in complete
%generality, even when the data take values in a countably
%infinite alphabet 
\cite{antos-K:01}.

A natural generalization is the problem of estimating 
the rate-distortion function 
$R(P,D)$ of a (not necessarily discrete-valued) source. 
Motivation for this comes in part
from lossy data compression, where we may need an estimate 
of how well a given data set could potentially be compressed,
cf.\ \cite{cover-wesel:94},
and also from cases where we want to quantify the
``information content'' of a particular signal,
but the data under examination take values 
in a continuous (or more general) alphabet,
cf.\ \cite{koval-rytsar:01}. 

The rate-distortion function estimation question 
appears to have received little attention in the
literature.  Here we present some basic results
for this problem.
% primarily focusing on consistency theorems.
% for a certain
% class of estimation problems.
% parametric and nonparametric estimators of $R(P,D)$.
First, we consider the simple {\bf plug-in estimator}
$R(P_{X_1^n},D)$, and determine conditions under
which it is strongly consistent, that is, it converges 
to $R(P,D)$ with probability 1, as $n\to\infty.$
We call this the {\bf nonparametric
estimation problem}, for reasons that will become clear
below. 

At first glance, consistency may seem to be a mere
continuity issue: Since the
empirical distribution $P_{X_1^n}$ converges,
with probability 1, to the true distribution 
$P$ as $n\to\infty$, a natural approach
to proving that $R(P_{X_1^n},D)$ also converges 
to $R(P,D)$ would be to try and establish some
sort of continuity property for $R(P,D)$ as a
function of $P$. But, as we shall see, 
$R(P_{X_1^n},D)$
turns out to be consistent under rather mild assumptions,
which are in fact {\em too mild} to ensure continuity
in any of the usual topologies;
see Section~\ref{ex:dc} for explicit
counterexamples.
This also explains our choice of the empirical distribution
$P_{x_1^n}$ as an estimate for $P$:
If $R(P,D)$ was continuous in $P$,
then any consistent estimator $\hat P_n$ of $P$ 
could be used to make $R(\hat P_n,D)$ 
a consistent estimator for $R(P,D)$.
% We do not further explore this continuity approach to consistency  
% because it seems to require rather restrictive assumptions and is 
% technically challenging.
Some of the subtleties in establishing regularity
properties of the rate-distortion function $R(P,D)$
as a function of $P$ are illustrated in
\cite{Csi:Extremum:1974}\cite{ahlswede:90}.

Another advantage of a plug-in estimator 
is that $P_{x_1^n}$ has finite 
support, regardless of the source alphabet.
This makes it possible (when the
reproduction alphabet is also finite)
to actually compute
$R(P_{x_1^n},D)$ by
approximation techniques
such as the
% For example, if 
% the reproduction alphabet is finite (or if it can be restricted to 
% a known finite set based on the data), then the 
Blahut-Arimoto algorithm
\cite{Bla:Computation:1972}\cite{Ari:Algorithm:1972}\cite{csiszar:74}.
% algorithm can be used.  
When the reproduction alphabet is continuous, % as well,
the Blahut-Arimoto algorithm can still be used after
discretizing the reproduction alphabet; the
discretization can,
in part, be justified by the observation that 
it can be viewed as an instance of the 
{\em parametric estimation problem} described below.
Other possibilities for continuous 
reproduction
alphabets are explored 
in \cite{Ros:Mapping:1994}\cite{Ben:Rate:1973}.
% Unfortunately, the 
% results here are only consistency results and do not address computation 
% or rates of convergence.  
% Rate of convergence results would be quite useful 
% for deciding whether or not the data requirements needed for accurate 
% estimation of $R(P,D)$ still permit feasible computation of its estimator.  
% Such results would likely require much stronger assumptions than do the 
% consistency results presented here.

The consistency problem can be framed 
in the following more general setting. 
As has been observed by several authors 
recently, the rate-distortion function 
of a memoryless source admits the 
decomposition,
\be
R(P,D)=\inf_Q R(P,Q,D),
\label{eq:decomp}
\ee
where the infimum is over all probability
distributions $Q$ on the reproduction alphabet,
and $R(P,Q,D)$ is the rate achieved by 
memoryless random codebooks with distribution $Q$
used to compress the source data to within distortion 
$D$; see, e.g., \cite{yang-kieffer:1}\cite{dembo-kontoyiannis:wyner}.
Therefore, $R(P,D)$ is the best rate that 
can be achieved by this family of codebooks.
But in the case where we only have a restricted family
of compression algorithms available, indexed,
say, by a family of probability distributions
$\{Q_\theta\,;\,\theta\in\Theta\}$ on the
reproduction alphabet, then the best achievable
rate is:
\be
R^\Theta(P,D):=\inf_{\theta\in\Theta} R(P,Q_\theta,D).
\label{eq:decomp2}
\ee
We also consider the 
{\bf parametric estimation problem},
namely, that of establishing the
strong consistency of the
corresponding {\bf plug-in estimator}
$R^\Theta(P_{X_1^n},D)$ as an estimator
for $R^\Theta(P,D)$.  It is important to
note that, when $\Theta$ indexes the set 
of all probability distributions on the 
reproduction alphabet, then the parametric and 
nonparametric problems are identical, and
this allows us to treat both problems in a common framework.

Our two main results, Theorems~\ref{t:ub} and~\ref{t:lb}
in the following section, give regularity conditions
for both the parametric
and nonparametric estimation problems
under which the plug-in estimator is strongly
consistent. It is shown that consistency 
holds in great generality for all distortion
values $D$ such that $R^\Theta(P,D)$ is 
continuous from the left. An example
illustrating that consistency may actually
fail at those points is given
in Section~\ref{ex:fail}. In particular,
for the nonparametric estimation problem we
obtain the following 
%three simple corollaries,
%which cover many of the interesting cases
%encountered in practice.
three simple corollaries, which cover many practical cases.

\vskip 1ex

\begin{cor} \label{c:finite} If the reproduction alphabet is finite, 
then for any source distribution $P$, $R(P_{X_1^n},D)$ is strongly 
consistent for $R(P,D)$ at all distortion levels $D\geq 0$ except perhaps 
at the single value where $R(P,D)$ 
transitions from being finite to being infinite.
\end{cor}  

\vskip 1ex

\begin{cor} \label{c:mse} If the source and reproduction
alphabets are both equal to
$\mathbb{R}^d$ and the distortion measure is 
squared-error,
then for any source distribution $P$
and any distortion level $D\geq 0$,
$R(P_{X_1^n},D)$ is strongly 
consistent for $R(P,D)$.
\end{cor}  

\vskip 1ex

\begin{cor} 
\label{c:sc} 
Assume that the reproduction alphabet is a compact, separable metric 
space, and that the distortion measure $\rho(x,\cdot)$ is continuous 
for each $x\in A$. Then (under mild additional measurability assumptions),
for any source distribution $P$,
$R(P_{X_1^n},D)$ is strongly 
consistent for $R(P,D)$ at all distortion levels $D\geq 0$ except perhaps 
at the single value where $R(P,D)$ 
transitions from being finite to being infinite.
\end{cor}

\vskip 1ex

Corollaries~\ref{c:finite} and~\ref{c:sc} are special cases of
Corollary~\ref{c:sc2} in Section~\ref{s:results}.   
Corollary~\ref{c:mse} is established
in Section~\ref{s:examples}, which
contains many other explicit examples
illustrating the consistency results and cases where 
consistency may fail. 
Section~\ref{s:proofs} contains
the proofs of all the main results in this paper.

% \redstart{Corollary~\ref{c:finite} is a special case of 
% Corollary~\ref{c:sc2} in Section~\ref{s:results} where the reproduction 
% alphabet is allowed to be any compact, separable metric space as long 
% as the distortion measure $\rho(x,\cdot)$ is continuous for each $x\in A$.  
% Corollary~\ref{c:mse} is established in Section~\ref{s:examples}, which
% contains many other explicit examples
% illustrating the consistency results and cases where 
% consistency may fail. 
% Section~\ref{s:proofs} contains
% the proofs of all the main results in this paper.}\redend

We also consider extensions of these results in two
directions. In Section~\ref{s:optimal} we examine 
the problem of estimating the optimal 
reproduction distribution -- namely, the distribution
that actually achieves the infimum in equations
(\ref{eq:decomp}) and (\ref{eq:decomp2}) -- from
empirical data. Consistency results are given, 
under conditions identical to those required
for the consistency of the plug-in estimator.
Finally, in Section~\ref{s:general} we show that
consistency holds for a more general class of estimators,
which arise as modifications of the plug-in.
These include, in particular, penalized versions
of the plug-in, analogous to the standard
penalized maximum likelihood estimators
used in statistics.

The analysis of the plug-in estimator presents
some unexpected technical difficulties. One way 
to explain the source of these difficulties
is by noting that there is a very close 
analogy, at least on the level of the 
mathematics, with the problem of maximum likelihood
estimation [see also Section~\ref{s:general} 
for another instance of this connection].
Beyond the superficial observation that
they are both extremization problems over a space
of probability distributions, a more accurate, albeit 
heuristic, illustration can be given as follows:
Suppose we have a memoryless source with distribution 
$P$ on some discrete alphabet, take the
reproduction alphabet to be the same as the source
alphabet,
and look at the extreme case where no 
% \redstart{(Hamming)}\redend 
distortion is allowed.
Then the plug-in estimator of the
rate-distortion function (which now is simply
the entropy) can be expressed as a 
trivial minimization over all possible coding 
distributions,
i.e., 
$$H(P_{x_1^n})=\min_{Q}[H(P_{x_1^n})+H(P_{x_1^n}\|Q)]
=-\frac{1}{n}\,\max_Q \big[\log Q^n(x_1^n)\big],$$
where $H(P\|Q)$ denotes the relative entropy,
and $Q^n$ is the $n$-fold
product 
distribution of $n$ independent random 
variables each distributed
according to $Q$.
Therefore, the computation of the plug-in estimate
$H(P_{x_1^n})$ is exactly equivalent to the computation
of the maximum likelihood estimate (MLE) of $P$ 
over a class of distributions $Q$. Alternatively,
% observe that, 
in Csisz\'ar's terminology,
the minimization of the relative entropy above
corresponds to the so-called ``reversed $I$-projection'' 
of $P_{x_1^n}$ onto the set of feasible distributions 
$Q$, which in this case consists of {\em all} distributions
on the reproduction alphabet; see, e.g.,
\cite{dykstra-lemke:88}\cite{csiszar-shields:04}.
Formally, this projection is
exactly the same as the computation of the MLE 
of $P$ based on $x_1^n$.

In the general case of nonzero distortion $D>0$,
the plug-in estimator can similarly be expressed
as, $R(P_{x_1^n},D)=\min_{Q}\,R(P_{x_1^n},Q,D)$,
cf.\ (\ref{eq:decomp}) above.
This (now highly nontrivial) minimization 
is mathematically very closely related to the 
problem of computing an $I$-projection as before.
The tools we employ to analyze this minimization
are based on the technique of epigraphical 
convergence \cite{Sal:Consistency:2001}\cite{AttWet:Epigraphical:1989}
(this is particularly clear in the proof 
of our main result, the lower bound 
in Theorem~\ref{t:lb}),
and it is no coincidence that these same tools
have also provided one of the most 
successful approaches to proving the
consistency of MLEs.
By the same token,
this connection 
also explains why the consistency
of the plug-in estimator involves 
subtleties similar to those cases where 
MLEs fail to be consistent \cite{lecam:90}.

In the way of motivation, we also mention 
that the asymptotic behavior of the plug-in estimator -- 
and the technical intricacies involved in its analysis --
also turn out to be important in extending some 
of Rissanen's celebrated ideas related 
to the Minimum Description Length (MDL) principle
to the context of lossy data compression;
this direction will be explored in subsequent work.

% Our main contribution here is a collection of 
% consistency results for the plug-in estimator 
% (and its parametric generalization),
% under completely
% general conditions. For example, 
% Two corollaries of our main results,
% which are easy to state and which cover many practical 
% situations, are the following:

Throughout the paper we work with stationary and ergodic sources 
instead of memoryless sources, though we are still only interested
in estimating the first-order rate-distortion function.
One reason for this is that the full rate-distortion
function can be estimated by looking at the process
in sliding blocks of length $m$ and then estimating the
``marginal'' rate-distortion function of these blocks
for large $m$;
see Section~\ref{ex:block}.
Another reason for allowing dependence in the data
comes from simulation:
For example, suppose we were interested in estimating 
the rate-distortion function of a distribution $P$ that 
we cannot compute explicitly (as is the case for perhaps
the majority of models used in image processing),
but for which we have a Markov chain Monte Carlo 
(MCMC) sampling algorithm.  The data generated by such 
an algorithm is not memoryless, yet we care only about 
the rate-distortion function of the marginal distribution.
In Section~\ref{s:LLN} we comment further on
this issue, and also give consistency results for data
produced by sources that may not be stationary.

%%% IK: later: expand discussion of main results, explain
%%% the non-parametric estimator, mention methods and techniques
%%% used in the analysis.

%%%%%%%%%%%%%%%%%%%%%%%%%%%%%%%%%%%%%%%%%%%%%%%%%
%%%%%%%%%%%%%%%%%%%%%%%%%%%%%%%%%%%%%%%%%%%%%%%%%
%%
% %  INTRODUCTION:   ORGANIZATION OF THE PAPER
%%
%%%%%%%%%%%%%%%%%%%%%%%%%%%%%%%%%%%%%%%%%%%%%%%%%
%%%%%%%%%%%%%%%%%%%%%%%%%%%%%%%%%%%%%%%%%%%%%%%%%

% \subsection{Organization of the paper}

%%%%%%%%%%%%%%%%%%%%%%%%%%%%%%%%%%%%%%%%%%%%%%%%%
%%%%%%%%%%%%%%%%%%%%%%%%%%%%%%%%%%%%%%%%%%%%%%%%%
%%
% %  MAIN RESULTS
%%
%%%%%%%%%%%%%%%%%%%%%%%%%%%%%%%%%%%%%%%%%%%%%%%%%
%%%%%%%%%%%%%%%%%%%%%%%%%%%%%%%%%%%%%%%%%%%%%%%%%

\section{Main Results}
\label{s:results}

%%%%%%%%%%%%%%%%%%%%%%%%%%%%%%%%%%%%%%%%%%%%%%%%%
%%%%%%%%%%%%%%%%%%%%%%%%%%%%%%%%%%%%%%%%%%%%%%%%%
%%
% %  MAIN RESULTS:   NOTATION
%%
%%%%%%%%%%%%%%%%%%%%%%%%%%%%%%%%%%%%%%%%%%%%%%%%%
%%%%%%%%%%%%%%%%%%%%%%%%%%%%%%%%%%%%%%%%%%%%%%%%%

% \subsection{Notation}

We begin with some notation and definitions that 
will remain in effect throughout the paper.

Suppose the random source $(X_n\,;\,n\geq 1)$ 
taking values in the source alphabet $A$ is to
be compressed in the reproduction alphabet
$\hat{A}$, with respect to the single-letter
distortion measures $(\rho_n)$ arising from
an arbitrary distortion function
$\rho:A\times\hat{A}\mapsto [0,\infty)$.
We assume that $A$ and $\hat{A}$ are equipped with the $\sigma$-algebras
${\cal A}$ and $\hat{\mathcal{A}}$, respectively,
that $(A,\mathcal{A})$ and $(\hat{A},\hat{\mathcal{A}})$ 
are Borel spaces, and that $\rho$ is 
$\sigma(\mathcal{A}\times\hat{\mathcal{A}})$-measurable.\footnote{Borel 
	spaces 
	include the Euclidean spaces ${\mathbb R}^d$ as well as all Polish 
	spaces, and they allow us to avoid certain measure-theoretic
	pathologies while working with 
	random sequences and conditional 
	distributions \cite{Kal:Foundations:2002}.
	Henceforth, all $\sigma$-algebras and the various product 
	$\sigma$-algebras derived from them are understood from the 
	context.  We do not complete any of 
	the $\sigma$-algebras, but we say that an event $C$ holds 
	{\it with probability 1} (w.p.1)~if $C$ contains a measurable 
	subset $C'$ 
	that has probability 1.} 
Suppose the source is stationary, and let $P$ denote 
its marginal distribution on $A$. Then
the (first-order) rate-distortion function $R_1(P,D)$ 
with respect to the distortion measure $\rho$ is defined as,
\[ R_1(P,D) := \inf_{(U,V)\sim W\in W(P,D)} I(U;V),
\;\;\;\;D\geq 0,\]
where the infimum is over all $A\times\hat{A}$-valued
random variables $(U,V)$ with joint distribution $W$
belonging to the set
\[ W(P,D) := \left\{W : W^A=P, \ E_W[\rho(U,V)] \leq D\right\} , \]
and where
$W^A$ denotes the marginal distribution of $W$ on $A$, 
and similarly for $W^{\hat{A}}$;
the infimum is taken to be
$+\infty$ when $W(P,D)$ is empty.
As usual, the mutual information 
$I(U;V)$
between two random variables $U,V$
with joint distribution $W$,
is defined as the relative entropy
between $W$ and the product of its
two marginals, $W^A\times W^{\hat{A}}$.
Here and throughout the paper, all 
familiar information-theoretic quantities
are expressed in nats, and $\log$ denotes
the natural logarithm. In particular,
for any two probability measures $\mu,\nu$
on the same space, the relative entropy
$H(\mu\|\nu)$ is
defined as $E_\mu [\log \frac{d\mu}{d\nu}]$
whenever the density $d\mu/d\nu$ exists, and 
it is taken to be $+\infty$ otherwise.
% \[ & \text{if $\mu\ll\nu$,} \\ \infty & \text{otherwise}. \end{cases} \]
% H(W\|\matsize{W^A}{W^{\hat{A}}}),  \]
% Note that $H(W\|\matsize{W^A}{W^{\hat{A}}})$ is simply 
% the mutual information $I(U;V)$ between two random variables $(U,V)$ 
% with joint distribution $W$.   

We write $\Dc(P)$ for
the set of distortion values $D\geq 0$
for which $R_1(P,D)$ 
is continuous from the left, i.e.,
\[ \Dc(P) := \left\{D \geq 0 : R_1(P,D) 
= {\textstyle \lim_{\lambda\uparrow 1}} R_1(P,\lambda D)\right\} . \]
By convention, this set always includes $0$ and any value of $D$ for which 
$R_1(P,D)=\infty$.  But since $R_1(P,D)$ is nonincreasing and convex in $D$ 
\cite{cover:book}\cite{Csi:Extremum:1974}, 
$\Dc(P)$ actually includes {\em all} $D\geq 0$ with the only
possible exception of the single value of $D$ where $R_1(P,D)$ 
transitions from being finite to being infinite.  
Conditions guaranteeing that $\Dc(P)$ is indeed 
all of $[0,\infty)$
can be found in \cite{Csi:Extremum:1974}.

% If $(X_n)_{n\geq 1}$ is a stationary memoryless source taking values in $A$
% and with marginal distribution $P$,
% then $R_1(P,D)$ is the rate-distortion function of the 
% source (with respect to the sequence of single-letter fidelity criteria
% based on $\rho$).   
% In the more general setting where
% $(X_n)_{n\geq 1}$ is not memoryless, but stationary and ergodic with 
% marginal distribution $P$, i.e., with $X_1\sim P$, then $R_1(P,D)$ is just 
% the first-order rate-distortion function of the source. 
  
%%%%%%%%%%%%%%%%%%%%%%%%%%%%%%%%%%%%%%%%%%%%%%%%%
%%%%%%%%%%%%%%%%%%%%%%%%%%%%%%%%%%%%%%%%%%%%%%%%%
%%
% %  MAIN RESULTS:   ESTIMATORS
%%
%%%%%%%%%%%%%%%%%%%%%%%%%%%%%%%%%%%%%%%%%%%%%%%%%
%%%%%%%%%%%%%%%%%%%%%%%%%%%%%%%%%%%%%%%%%%%%%%%%%

\subsection{Estimation Problems and Plug-in Estimators}

Given a finite-length data string
$x_1^n := (x_1,x_2,\dotsc,x_n)$ produced 
by a stationary source $(X_n)$ as above
with marginal distribution $P$,
the {\bf plug-in estimator} of the first-order
rate-distortion function $R_1(P,D)$ is
$R_1(P_{x_1^n},D)$, where $P_{x_1^n}$ is
the {\em empirical distribution}
induced by the sample $x_1^n$ on $A^n$, namely,
\[ P_{x_1^n}(C) := \frac{1}{n}\sum_{k=1}^n 
\ind\{x_k \in C\}  \quad \quad x_1^n\in A^n , \ C \in \mathcal{A} \]
and where $\ind$ is the indicator function.
Our first goal is to obtain conditions under
which this estimator is strongly consistent.
We call this the {\bf nonparametric estimation problem}.

We also consider the more general class of estimation
problems mentioned in the Introduction.
Suppose for a moment that our goal is to compress data 
produced by a {\em memoryless} source $(X_n)$ 
with distribution $P$ on $A$, and suppose also
that we are restricted to using memoryless random 
codebooks with distributions $Q$ belonging
to some parametric family $\{Q_\theta\;:\;\theta\in\Theta\}$ 
where $\Theta$
indexes a subset of all probability distributions on 
$\hat{A}$. Using a random codebook with distribution $Q$
to compress the data to within distortion $D$,
yields (asymptotically) a rate of $R_1(P,Q,D)$ 
nats/symbol, where the rate-function $R_1(P,Q,D)$
is given by,
$$
R_1(P,Q,D)=
\inf_{W\in W(P,D)} H(W\|\matsize{P}{Q}).
$$
See \cite{yang-kieffer:1}\cite{dembo-kontoyiannis:wyner}
for details. From this it is immediate that 
the rate-distortion function
of the source admits the decomposition
given in (\ref{eq:decomp}).
% \[ R_1(P,D) = \inf_Q  
% R_1(P,Q,D) , \]
% where the infimum is over all probability
% distributions $Q$ on $\hat{A}.$
Having restricted attention
to the class of codebook distributions
$\{Q_\theta\;;\;\theta\in\Theta\},$ then the best 
possible compression rate is:
\begin{equation} 
R_1^\Theta(P,D) := \inf_{\theta\in\Theta} R_1(P,Q_\theta,D)
\;\;\mbox{nats/symbol.}
\label{e:R} 
\end{equation}
When $\theta$ indexes certain nice families, 
say Gaussian, the infimum $R_1^\Theta(P,D)$ can be analytically 
derived or easily computed, often for any distribution $P$, 
including an empirical distribution.  

Thus motivated, we now formally define the
{\bf parametric estimation problem}.
Suppose $(X_n)$ is a stationary source
as above, and let $\{Q_\theta\;:\;\theta\in\Theta\}$ be a family 
of probability distributions on the reproduction alphabet $\hat{A}$
parameterized by an arbitrary parameter space $\Theta$.
The {\bf plug-in estimator} for 
$R_1^\Theta(P,D)$ is 
$R_1^\Theta(P_{X_1^n},D)$, and we seek conditions
for its strong consistency.

Note that $R_1^\Theta(P,D)=R_1(P,D)$
when $\{Q_\theta\;:\;\theta\in\Theta\}$ includes 
all probability distributions on $\hat{A},$
or if it simply includes the optimal reproduction
distribution achieving the infimum in (\ref{eq:decomp}).
Otherwise, 
$R_1^\Theta(P,D)$ may be strictly larger 
than $R_1(P,D)$. Therefore, the nonparametric
problem is a special case of the parametric one,
and we can consider the two situations 
in a common framework.

In the parametric scenario we write,
\[ \Dc^\Theta(P) := \left\{D \geq 0 : R_1^\Theta(P,D) = 
{\textstyle \lim_{\lambda\uparrow 1}} R_1^\Theta(P,\lambda D)\right\} . \]
Unlike $\Dc(P)$, $\Dc^\Theta(P)$ can exclude 
more than a single point.

%%%%%%%%%%%%%%%%%%%%%%%%%%%%%%%%%%%%%%%%%%%%%%%%%
%%%%%%%%%%%%%%%%%%%%%%%%%%%%%%%%%%%%%%%%%%%%%%%%%
%%
% %  MAIN RESULTS:   STATISTICAL CONSISTENCY
%%
%%%%%%%%%%%%%%%%%%%%%%%%%%%%%%%%%%%%%%%%%%%%%%%%%
%%%%%%%%%%%%%%%%%%%%%%%%%%%%%%%%%%%%%%%%%%%%%%%%%

\subsection{Consistency}

We investigate conditions under which the 
plug-in estimator $R^\Theta_1(P_{x_1^n},D)$
is strongly consistent, i.e.,\footnote{Throughout the paper 
	we do not require limits to be finite valued, but say that 
	$\lim_n a_n = \infty$ if $a_n$ diverges to $\infty$ 
	(and similarly for $-\infty$).}
\begin{equation} 
R_1^\Theta(P_{X_1^n},D)\relas{\to} R_1^\Theta(P,D). 
\label{e:sc} 
\end{equation}  
Of course in the special case where 
$\Theta$ indexes all probability distributions on $\hat{A}$, 
this reduces to the nonparametric problem, and (\ref{e:sc})
becomes $R_1(P_{X_1^n},D)\relas{\to} R_1(P,D)$.
We separately treat the upper and lower bounds 
that combine to give \eqref{e:sc}.

%%%%%%%%%%%%%%%%%%%%%%%%%%%%%%%%%%%%%%%%%%%%%%%%%
%%%%%%%%%%%%%%%%%%%%%%%%%%%%%%%%%%%%%%%%%%%%%%%%%
%%
% %  MAIN RESULTS:   STATISTICAL CONSISTENCY:   UPPER BOUND
%%
%%%%%%%%%%%%%%%%%%%%%%%%%%%%%%%%%%%%%%%%%%%%%%%%%
%%%%%%%%%%%%%%%%%%%%%%%%%%%%%%%%%%%%%%%%%%%%%%%%%

% \subsubsection{Upper Bound}

The upper bound does not require {\em any} further regularity 
assumptions, although there can be certain pathological values 
of $D$ for which it is not valid.  
In the nonparametric situation, the only potential problem point is the 
single value of $D$ where $R_1(P,D)$ transitions from finite to infinite.

\vskip 1ex
\begin{thm} \label{t:ub} If the source $(X_n)$ is stationary and ergodic 
with $X_1\sim P$, then
\[ \limsup_{n\to\infty} R_1^\Theta(P_{X_1^n},D) \relas{\leq} R_1^\Theta(P,D) \]
for all $D\in\Dc^\Theta(P)$. 
\end{thm}
\vskip 1ex

As illustrated by a simple
counterexample in Section~\ref{ex:fail},
the requirement that $D\in\Dc^\Theta(P)$ 
cannot be relaxed completely.
The proof of the theorem, given in Section~\ref{s:proofs},
is a combination of the decomposition in \eqref{e:R} and the 
fact that $R_1(P_{X_1^n},Q,D)\relas{\to} R_1(P,Q,D)$ quite generally.
Actually, from the proof we also obtain 
% \redstart{stronger}\redend %weaker 
an upper bound on the $\liminf$,
\begin{equation} 
\liminf_{n\to\infty} R_1^\Theta(P_{X_1^n},D) \relas{\leq} R_1^\Theta(P,D) \ \ 
\text{for all $D\geq 0$}, 
\label{e:liminf} 
\end{equation}
which 
provides some information even for those values of $D$ 
where the upper bound in Theorem \ref{t:ub} may fail.  

%%%%%%%%%%%%%%%%%%%%%%%%%%%%%%%%%%%%%%%%%%%%%%%%%
%%%%%%%%%%%%%%%%%%%%%%%%%%%%%%%%%%%%%%%%%%%%%%%%%
%%
% %  MAIN RESULTS:    STATISTICAL CONSISTENCY:   LOWER BOUND
%%
%%%%%%%%%%%%%%%%%%%%%%%%%%%%%%%%%%%%%%%%%%%%%%%%%
%%%%%%%%%%%%%%%%%%%%%%%%%%%%%%%%%%%%%%%%%%%%%%%%%

% \subsubsection{Lower Bound}

For the corresponding lower bound in \eqref{e:sc},
some mild additional assumptions are needed.
We will always assume that $\Theta$ 
is a metric space, and also that the following
two conditions are satisfied:

\begin{itemize}
\item[A1.] The map $\theta\mapsto E_\theta [e^{\lambda\rho(x,Y)}]$ 
is continuous for each $x\in A$ and $\lambda\leq 0$,
where $E_\theta$ denotes expectation w.r.t.~$Q_\theta$.  
\item[A2.] For each $D\geq 0$, there exists a (possibly random) sequence 
$(\theta_n)$ with
\begin{equation} 
\label{e:A2} 
\liminf_{n\to\infty} R_1(P_{X_1^n},Q_{\theta_n},D) 
\relas{\leq} \liminf_{n\to\infty} R_1^\Theta(P_{X_1^n},D),
\end{equation}
and such that $(\theta_n)$ is relatively compact with
probability 1.
\end{itemize}

\vskip 1ex
\begin{thm} \label{t:lb} If $\Theta$ is separable, A1 and A2 hold, 
and $(X_n)$ is stationary and ergodic with $X_1\sim P$,
then
\[ \liminf_{n\to\infty} R_1^\Theta(P_{X_1^n},D)  
\relas{\geq} R_1^\Theta(P,D) \] 
for all $D\geq 0$. 
\end{thm}
\vskip 1ex

Although A1 and A2 may seem quite involved,
they are fairly easy to verify in specific
examples:
For A1, we have the following sufficient conditions;
as we prove in Section~\ref{s:proofs},
either one implies A1.
\begin{itemize}
\item[P1.] Whenever $\theta_n\to\theta$,
we also have that $Q_{\theta_n}\to Q_\theta$ setwise.\footnote{We 
say that $Q_m\to Q$ setwise if $E_{Q_m} (f)\to E_Q (f)$ for all bounded, 
measurable functions $f$, or equivalently, if $Q_m(C)\to Q(C)$ 
for all measurable sets $C$.}  
\item[N1.] $(\hat{A},\hat{\mathcal{A}})$ is a metric space with its Borel 
$\sigma$-algebra, $\rho(x,\cdot)$ is continuous for each $x\in A$ and 
$\theta_n\to\theta$ implies that $Q_{\theta_n}\to Q_\theta$ 
weakly.\footnote{We say that $Q_m\to Q$ weakly if $E_{Q_m} (f)\to E_Q (f)$ 
for all bounded, continuous functions $f$, or equivalently, 
if $Q_m(C)\to Q(C)$ for all measurable sets $C$ with $Q(\partial C)=0$.}  
\end{itemize}

For A2, we first note that 
a sequence $(\theta_n)$ satisfying \eqref{e:A2} 
{\em always} exists and 
that the inequality in \eqref{e:A2} must always be an equality.
The important requirement in A2 is that $(\theta_n)$ 
be relatively compact.  In particular, A2 is trivially true 
if $\Theta$ is compact.  
More generally, the following two conditions make 
it easier to verify A2 in particular examples. 
In Section~\ref{s:proofs}
we prove that either one implies A2 as long as the source 
is stationary and ergodic with marginal distribution $P$.
For any subset $K$ of the
source alphabet $A$, we write $B(K,M)$ for the
subset of $\hat{A}$ which is the union of all the
distortion balls of radius $M\geq 0$ centered at
points of $K$. Formally,
\[ B(K,M) := \bigcup_{x\in K} \{y : \rho(x,y) \leq M\} , 
\quad K\subseteq A, \ M \geq 0 . \]

\begin{itemize}
\item[P2.] For each $D\geq 0$, there exists a $\Delta > 0$ and a 
$K\in\mathcal{A}$ such that $P(K) > D/(D+\Delta)$ and 
$\{\theta : Q_\theta(B(K,D+\Delta)) \geq \epsilon\}$ is relatively 
compact for each $\epsilon > 0$.
\item[N2.] $(\hat{A},\hat{\mathcal{A}})$ is a metric space with its 
Borel $\sigma$-algebra, $\Theta$ is the set of all probability distributions 
on $\hat{A}$ with a metric that metrizes weak 
convergence of probability measures, and for each $\epsilon>0$ and each 
$M>0$ there exists a $K\in\mathcal{A}$ such that $P(K) > 1-\epsilon$ and 
$B(K,M)$ is relatively compact.\footnote{$\Theta$ can always be 
metrized in this way, and so that $\Theta$ will be separable 
(compact) if $\hat{A}$ is separable (compact) \cite{billingsley:cpm}. 
\label{f:np}}
\end{itemize}

In Section~\ref{s:examples} we describe concrete situations 
where these assumptions are valid.   

The proof of Theorem~\ref{t:lb} has the following main
ingredients. The separability of $\Theta$ and the continuity 
in A1 are used to ensure measurability and, in particular,
for controlling exceptional sets. A1 is a local assumption 
that ensures $\inf_{\theta\in U} R_1(P_{X_1^n},Q_\theta,D)$ 
is well behaved in small neighborhoods $U$.  A2 is a global 
assumption that ensures the final analysis can be restricted 
to a small neighborhood. 
% Notice that \eqref{e:liminf} implies 
% that the inequality 
% in Theorem~\ref{t:lb} is actually an equality.

Combining Theorems~\ref{t:ub} and~\ref{t:lb} gives conditions under 
which $R_1^\Theta(P_{X_1^n},D)\relas{\to} R_1^\Theta(P,D)$.  
In the nonparametric situation we have the following Corollary,
which is a generalization of 
% \redstart{Corollary~\ref{c:finite}}\redend
Corollary~\ref{c:sc} 
in the Introduction;
it follows immediately from the last two theorems.

\vskip 1ex
\begin{cor} 
\label{c:sc2} 
Suppose $(\hat{A},\hat{\mathcal{A}})$ is a compact, separable metric space 
with its Borel $\sigma$-algebra and $\rho(x,\cdot)$ is continuous for each 
$x\in A$.  If $(X_n)$ is stationary and ergodic with $X_1\sim P$, 
then $R_1(P_{X_1^n},D) \relas{\to} R_1(P,D)$ for all $D\in\Dc(P)$.
Furthermore, the compactness condition can be relaxed as in N2.
\end{cor}

%%%%%%%%%%%%%%%%%%%%%%%%%%%%%%%%%%%%%%%%%%%%%%%%%
%%%%%%%%%%%%%%%%%%%%%%%%%%%%%%%%%%%%%%%%%%%%%%%%%
%%
% %  EXAMPLES
%%
%%%%%%%%%%%%%%%%%%%%%%%%%%%%%%%%%%%%%%%%%%%%%%%%%
%%%%%%%%%%%%%%%%%%%%%%%%%%%%%%%%%%%%%%%%%%%%%%%%%

\section{Examples} \label{s:examples}

In all of the examples we assume that the source $(X_n)$ is 
stationary and ergodic with $X_1\sim P$.

%%%%%%%%%%%%%%%%%%%%%%%%%%%%%%%%%%%%%%%%%%%%%%%%%
%%%%%%%%%%%%%%%%%%%%%%%%%%%%%%%%%%%%%%%%%%%%%%%%%
%%
% %  EXAMPLES:    NONPARAMETRIC DISCRETE
%%
%%%%%%%%%%%%%%%%%%%%%%%%%%%%%%%%%%%%%%%%%%%%%%%%%
%%%%%%%%%%%%%%%%%%%%%%%%%%%%%%%%%%%%%%%%%%%%%%%%%

\subsection{Nonparametric Consistency: Discrete Alphabets} 
\label{ex:discrete}

Let $A$ and $\hat{A}$ be at most countable and let $\rho$ be unbounded in 
the sense that for each fixed $x\in A$ and each fixed $M > 0$ there are 
only finitely many $y\in\hat A$ with $\rho(x,y) < M$.   N1 and N2 are 
clearly satisfied in the nonparametric setting where $\Theta$ is the 
set of all probability distributions on $\hat{A}$, so 
$R_1(P_{X_1^n},D)\relas{\to} R_1(P,D)$ for all $D$ except 
perhaps at the single value of $D$ where $R_1(P,D)$ transitions from 
finite to infinite. 
If, in addition, for each $x$ there exists a $y$ 
with $\rho(x,y)=0$, then $\Dc(P)=[0,\infty)$ regardless 
of $P$ \cite{Csi:Extremum:1974}, and the plug-in 
estimator is strongly consistent for all $P$ and all $D$.

This example also yields a different proof
of the general consistency result mentioned in the 
Introduction, for the plug-in estimate of the entropy
of a discrete-valued source:
%If we take $A=\hat{A}$ to be at most countable,
% \redstart{
If we map $A=\hat{A}$ into the integers,
% }\redend
let $\rho(x,y)=|x-y|$, and take $D=0$, then we 
obtain the strong consistency 
of \cite[Cor.~1]{antos-K:01}.

%%%%%%%%%%%%%%%%%%%%%%%%%%%%%%%%%%%%%%%%%%%%%%%%%
%%%%%%%%%%%%%%%%%%%%%%%%%%%%%%%%%%%%%%%%%%%%%%%%%
%%
% %  EXAMPLES:   NONPARAMETRIC CONTINUOUS
%%
%%%%%%%%%%%%%%%%%%%%%%%%%%%%%%%%%%%%%%%%%%%%%%%%%
%%%%%%%%%%%%%%%%%%%%%%%%%%%%%%%%%%%%%%%%%%%%%%%%%

\subsection{Nonparametric Consistency: Continuous Alphabets} \label{ex:np}  

Again in the nonparametric setting, let $A=\hat{A}=\mathbb{R}^d$ 
be finite dimensional Euclidean space, and let 
$\rho(x,y):= f(\|x-y\|)$ for some function $f$ of Euclidean distance where 
$f:[0,\infty)\to[0,\infty)$ is continuous and $f(t)\to\infty$ as 
$t\to\infty$.  As in the previous example, N1 and N2 are clearly satisfied, so $R_1(P_{X_1^n},D)\relas{\to} R_1(P,D)$ for all $D$ except 
perhaps at the single value of $D$ where $R_1(P,D)$ transitions from 
finite to infinite.  If furthermore $f(0)=0$, then $\Dc(P)=[0,\infty)$ 
regardless of $P$ \cite{Csi:Extremum:1974} and the
plug-in estimator is strongly consistent for all $P$ and all $D$.
 
This example includes the important special case of squared-error distortion: 
In the nonparametric problem, 
the plug-in estimator is always strongly consistent under 
squared-error distortion over finite dimensional Euclidean space,
as stated in Corollary~\ref{c:mse}.  This 
example also generalizes as follows.  The alphabets $A$ and $\hat{A}$ 
can be (perhaps different) subsets of $\mathbb{R}^d$, as long as 
$\hat{A}$ is closed.  The use of Euclidean distance is not essential 
and we can take any $\rho \geq f$, so that $\rho$ is not required to 
be translation invariant, as long as $\rho$ is continuous over 
$\hat{A}$ for each fixed $x\in A$.  This is enough for consistency 
except perhaps at a single value of $D$.  To use the results 
in \cite{Csi:Extremum:1974} to rule out any pathological values of $D$, 
that is, to show that $\Dc=[0,\infty)$ we also need $A$ to be closed, 
$\rho$ to be continuous over $A$ for each fixed $y$ 
and $\inf_y \rho(x,y) =0$ for each $x$.

%%%%%%%%%%%%%%%%%%%%%%%%%%%%%%%%%%%%%%%%%%%%%%%%%
%%%%%%%%%%%%%%%%%%%%%%%%%%%%%%%%%%%%%%%%%%%%%%%%%
%%
% %  EXAMPLES:   PARAMETRIC GAUSSIAN
%%
%%%%%%%%%%%%%%%%%%%%%%%%%%%%%%%%%%%%%%%%%%%%%%%%%
%%%%%%%%%%%%%%%%%%%%%%%%%%%%%%%%%%%%%%%%%%%%%%%%%

\subsection{Parametric Consistency for Gaussian Families} \label{ex:gaussian} 

Let $A=\hat{A}=\mathbb{R}$, let $\rho$ satisfy the 
assumptions of Example \ref{ex:np}, let 
$\Theta=\{(\mu,\sigma)\in\mathbb{R}\times[0,\infty)\}$ with the Euclidean 
metric, and for each $\theta=(\mu,\sigma)$ let $Q_\theta$ be Gaussian with 
mean $\mu$ and standard deviation $\sigma$ [the case $\sigma=0$ 
corresponds to 
the point mass at $\mu$].  Conditions N1 and P2 are clearly satisfied, so 
$R_1^\Theta(P_{X_1^n},D)\relas{\to}R_1^\Theta(P,D)$ for all 
$D\in\Dc^\Theta(P)$.  In the special case where $\rho(x,y) = (x-y)^2$ is 
squared-error distortion, then it is not too difficult 
\cite{dembo-kontoyiannis:wyner}
to show that
\[ R_1^\Theta(P,D) = \max\Big\{0,\frac{1}{2}\log\frac{\sigma_X^2}{D}\Big\} , \]
where $\sigma_X^2$ denotes the (possibly infinite) variance of $P$, so 
$\Dc^\Theta(P)=[0,\infty)$ and the convergence holds for all $D$. 
Furthermore, if the source $P$ happens to also be Gaussian, then 
$R_1^\Theta(P,D)=R_1(P,D)$ and
the plug-in estimator is also strongly consistent for the
nonparametric problem.

%%%%%%%%%%%%%%%%%%%%%%%%%%%%%%%%%%%%%%%%%%%%%%%%%
%%%%%%%%%%%%%%%%%%%%%%%%%%%%%%%%%%%%%%%%%%%%%%%%%
%%
% %  EXAMPLES:   CONVERGENCE FAILURE
%%
%%%%%%%%%%%%%%%%%%%%%%%%%%%%%%%%%%%%%%%%%%%%%%%%%
%%%%%%%%%%%%%%%%%%%%%%%%%%%%%%%%%%%%%%%%%%%%%%%%%

\subsection{Convergence Failure for $D\not\in\Dc(P)$} \label{ex:fail} 

Let $A=\{0,1\}$, $\hat{A}=\{0\}$, and $\rho(x,y):=|x-y|$.   
Since there is only one 
possible distribution on $\hat{A}$, it is easy to show that
\[ R_1(P',D) = \begin{cases} 0 & \text{if $P'(1) \leq D$} \\ \infty & 
\text{otherwise } \end{cases} \]
for any distribution $P'$ on $A$.  If $P(1) > 0$, the only possible 
trouble point for consistency is $D=P(1)$, which is not in $\Dc(P)$. 
It is easy to see that convergence (and therefore consistency) might fail at 
this point because $R_1(P_{X_1^n},D)$ will jump back and forth between $0$ 
and $\infty$ as $P_{X_1^n}(1)$ jumps above and below $D=P(1)$. 
The law of the iterated logarithm implies that this failure to converge 
happens with probability 1 when the source is memoryless.  In general, when the 
source is stationary and ergodic, it turns out that 
convergence will fail 
with positive probability 
\cite{Har:Gen:Sub}\cite{Har:Generalized:arXiv}\cite{Har:First:2003}.

%%%%%%%%%%%%%%%%%%%%%%%%%%%%%%%%%%%%%%%%%%%%%%%%%
%%%%%%%%%%%%%%%%%%%%%%%%%%%%%%%%%%%%%%%%%%%%%%%%%
%%
% %  EXAMPLES:   DISCONTINUITY
%%
%%%%%%%%%%%%%%%%%%%%%%%%%%%%%%%%%%%%%%%%%%%%%%%%%
%%%%%%%%%%%%%%%%%%%%%%%%%%%%%%%%%%%%%%%%%%%%%%%%%

\subsection{Consistency at a Point of Discontinuity in $P$} \label{ex:dc}

This slightly modified example from Csisz\'ar \cite{Csi:Extremum:1974} 
illustrates that $R_1(\cdot,D)$ can be discontinuous at $P$ even though 
the plug-in estimator is consistent.  Let $A=\hat{A}=\{1,2,\dotsc\}$, 
let $P'$ be any distribution on $A$ with infinite entropy and with $P'(x)>0$ 
for all $x$, and let $\rho(x,y) := P'(x)^{-1}\ind\{x\neq y\}+|x-y|$.   
Note that $R_1(P',D) = \infty$ for 
all $D$.\footnote{$R_1(P',\cdot)\equiv\infty$, 
	because any pair of random variables $(U,V)$ with $U\sim P'$ and 
	$E[\rho(U,V)] < \infty$ has $I(U;V)=\infty$.  To see this, 
	first note that  $E[\rho(U,V)] < \infty$ implies that 
	$\alpha(x):=\Prob\{V=x|U=x\}\to 1$ as $x\to\infty$;
	simply 
	use the definition of $\rho$ and ignore the $|x-y|$ term.
	Computing the mutual information and using the log-sum inequality 
	gives 
	$I(U;V) \geq \kappa +
	\sum_x P'(x)\alpha(x)\log\bigl(\alpha(x)/Q(x)\bigr)$, 
	where $V\sim Q$ and where $\kappa$ is a finite constant that 
	comes from all of the other terms in the definition of $I(U;V)$ 
	combined together with the log-sum inequality.  Since $\alpha(x)\to 1$
	and since $\sum_x P'(x)\log \bigl(1/Q(x)\bigr) \geq H(P')=\infty$ 
	for any probability distribution $Q$, we see that 
	$I(U;V)=\infty$.  We can ignore $\alpha(x)$ because the finiteness 
	of the sum only depends on the behavior for large $x$,
	and for large enough $x$ we have
	$\alpha(x) > 1/2$, say.}    
This is a special case of Example~\ref{ex:discrete} so the 
plug-in estimator is always strongly consistent regardless of $P$ and $D$.  
Nevertheless, $R_1(\cdot,D)$ is discontinuous everywhere it is finite. 

To see this, let the source $P$ be any distribution on $A$ with finite entropy 
$H(P)$.  Note that $R_1(P,D) \leq R_1(P,0) = H(P) < \infty$.  Define the 
mixture distribution $P_\epsilon := (1-\epsilon)P+\epsilon P'$.  Then 
$P_\epsilon \to P$ in the topology of total variation\footnote{The topology 
	of total variation is metrized by the distance 
	$d(P,P'):= \sup_{C}|P(C)-P'(C)|$.}  
(and also any weaker topology) as $\epsilon\downarrow 0$, but 
$R_1(P_\epsilon,D)\not\to R_1(P,D)$ because 
$R_1(P_\epsilon,D)\geq \epsilon R_1(P',D/\epsilon) = \infty$ for all 
$\epsilon > 0$.  See \eqref{e:Reps} below for a proof of this last 
inequality.\footnote{An interesting special case 
	of this example (based on the fact that $\sum_{x\geq 2} 
	[x\log^\alpha x]^{-1}$ converges if and only if $\alpha > 1$) is 
	% \cite[Thm. 3.29]{Rud:Principles:1976}
	$P'(x-1)\propto 1/(x\log^{1.5} x)$ (infinite entropy) and 
	$P(x-1)\propto 1/(x \log^{2.5} x)$ (finite entropy), 
	$x=2,3,\dotsc$, because the relative entropies $H(P'\|P)$ and 
	$H(P\|P')$ are both finite, so $H(P_\epsilon\|P)\to 0$ and 
	$H(P\|P_\epsilon)\to 0$ as $\epsilon\downarrow 0$.  (From the 
	convexity of relative entropy.) This counterexample thus shows
	that even closeness in relative entropy between two distributions
	(which is stronger than closeness in total variation) 
	is not enough to guarantee the closeness of the rate-distortion 
	functions of the corresponding distributions.} 

The key property of $\rho$ in this example is that there exists 
a $P'$ with $R_1(P',D)=\infty$ for {\em all} $D$.  If such a $P'$ exists, 
then $R_1(\cdot,D)$ will be discontinuous in the topology of total variation 
at any point $P$ where $R_1(P,D)$ is finite for exactly the same reason as 
above.  Although this specific example is based on a rather pathological 
distortion measure, many unbounded distortion measures on continuous 
alphabets, including squared-error distortion on $\mathbb{R}$, have such 
a $P'$ and are thus discontinuous in the topology of total 
variation.\footnote{For squared-error 
	distortion, let $P'$ be any distribution over discrete points 
	$\{x_1,x_2,\dotsc\}\subset\mathbb{R}$ where 
	$x_k \geq x_{k-1}+2^{1/P'(x_k)}$ and where $H(P')=\infty$.  
	This is essentially the same as Csisz\'ar's example above because 
	any pair of random variables $(U,V)$ with 
	$E[\rho(U,V)] < \infty$ must have 
	$\Prob\{V \text{ closer to $x_k$ than any other $x_j$}|U =x_k\}\to 1$
	as $k\to\infty$.}   

%%%%%%%%%%%%%%%%%%%%%%%%%%%%%%%%%%%%%%%%%%%%%%%%%
%%%%%%%%%%%%%%%%%%%%%%%%%%%%%%%%%%%%%%%%%%%%%%%%%
%%
% %  EXAMPLES:   HIGHER ORDER 
%%
%%%%%%%%%%%%%%%%%%%%%%%%%%%%%%%%%%%%%%%%%%%%%%%%%
%%%%%%%%%%%%%%%%%%%%%%%%%%%%%%%%%%%%%%%%%%%%%%%%%

\subsection{Higher-Order Rate-Distortion Functions} \label{ex:block}

Suppose that we want to estimate the $m$th-order rate-distortion function 
of a stationary and ergodic source $(X_n)$ with $m$th order 
marginal distribution $X_1^m\sim P_m$, 
namely,
\[ R_m(P_m,D) := \frac{1}{m} \inf_{(U,V)\sim W\in W_m(P_m,D)} I(U;V) , \]
where the infimum is over all $A^m\times\hat{A}^m$-valued random variables, 
with joint distribution $W$ in the set $W_m(P_m,D)$ of probability 
distributions on $A^m\times \hat{A}^m$ whose marginal distribution
on $A^m$ equals $P_m$, and which have 
$E[\rho_m(U,V)] \leq D$ for 
\[ \rho_m(x_1^n,y_1^n) := \frac{1}{m}\sum_{k=1}^m 
\rho(x_k,y_k) \quad \quad x_1^m\in A^m, \ y_1^m\in \hat{A}^m . \] 
All our results above immediately apply to this situation.  We 
simply estimate the first-order rate-distortion function of the 
sliding-block process $(Z_n)$ defined by $Z_k := (X_k,\dotsc,X_{k+m-1})$ 
with source alphabet $A^m$, reproduction alphabet $\hat{A}^m$ and 
distortion measure $\rho_m$, and then divide the estimate by $m$.

%%%%%%%%%%%%%%%%%%%%%%%%%%%%%%%%%%%%%%%%%%%%%%%%%
%%%%%%%%%%%%%%%%%%%%%%%%%%%%%%%%%%%%%%%%%%%%%%%%%
%%
% %  REMARKS AND EXTENSIONS
%%
%%%%%%%%%%%%%%%%%%%%%%%%%%%%%%%%%%%%%%%%%%%%%%%%%
%%%%%%%%%%%%%%%%%%%%%%%%%%%%%%%%%%%%%%%%%%%%%%%%%

\section{Further Results} \label{s:remarks}

%%%%%%%%%%%%%%%%%%%%%%%%%%%%%%%%%%%%%%%%%%%%%%%%%
%%%%%%%%%%%%%%%%%%%%%%%%%%%%%%%%%%%%%%%%%%%%%%%%%
%%
% %  REMARKS AND EXTENSIONS:   MINIMIZERS
%%
%%%%%%%%%%%%%%%%%%%%%%%%%%%%%%%%%%%%%%%%%%%%%%%%%
%%%%%%%%%%%%%%%%%%%%%%%%%%%%%%%%%%%%%%%%%%%%%%%%%

\subsection{Estimation of the Optimal Reproduction Distribution}
\label{s:optimal}

So far, we concentrated on conditions under which the
plug-in estimator is consistent; these guarantee
an (asymptotically) accurate estimate of the
best compression rate
$R_1^\Theta(P,D)=\inf_{\theta\in\Theta}R_1(P,Q_\theta,D)$
that can be achieved by codes
restricted to some class of distributions 
$\{Q_\theta\;;\;\theta\in\Theta\}$.
Now suppose this infimum 
is achieved by some $\theta^*$, 
corresponding to the optimal reproduction
distribution $Q_{\theta^*}$.
Here we use a simple modification of 
the plug-in estimator in order to obtain estimates
$\theta_n=\theta_n(x_1^n)$ for the optimal 
reproduction parameter $\theta^*$
based on the data $x_1^n$.
Specifically, since we have conditions
under which 
\be
\inf_{\theta\in\Theta}R_1(P_{x_1^n},Q_\theta,D)
\approx
\inf_{\theta\in\Theta}R_1(P,Q_\theta,D),
\label{eq:approx}
\ee
% \redstart{
we naturally consider the sequence of estimators
which achieve the infima on the left-hand-side of \eqref{eq:approx} 
for each $n\geq 1$; that is,  we simply replace the $\inf$
by an $\arginf$. Since these arg-infima may not exist or may not be unique, 
we actually consider any sequence of {\bf approximate minimizers} 
$(\theta_n)$ that have
$R_1(P_{X_1^n},Q_{\theta_n},D)\approx R_1^\Theta(P_{X_1^n},D)$
in the sense that \eqref{e:arginf} below holds.  
Similarly, minimizers $\theta^*$ of the right-hand-side of 
\eqref{eq:approx} may not exist or be unique, either.  
We thus consider the (possibly empty) set $\Theta^*$ containing 
all the minimizers of $R_1(P,Q_\theta,D)$ and address the problem 
of whether the estimators $\theta_n$ converge
%to some $\theta^*\in\Theta^*$.
to $\Theta^*$, meaning that $\theta_n$ is eventually in any neighborhood of $\Theta^*$.

%we naturally define the {\bf approximate
%minimizers} $(\theta_n)$ as the values
%which achieve the infima on the left-hand-side
%of (\ref{eq:approx}), for all $n\geq 1$.
%Since these infima may not be achieved 
%uniquely (or even at all), we consider the set 
%$\Theta^*$ containing all the minimizers of $R_1(P,Q_\theta,D)$,
%and we call $(\theta_n)$ a sequence of approximate minimizers
%for the data string $x_1^\infty=(x_1,x_2,\ldots)$, if
%$R_1(P_{x_1^n},Q_{\theta_n},D)=R_1^\Theta(P_{x_1^n},D)$
%for all large enough $n$, or, more generally,
%if \eqref{e:arginf} below holds. 

% i.e.,
% $$R_1(P_{X_1^n},D)=\inf_Q R_1(P_{x_1^n},D)
% \relas{\to}
% R_1(P)=\inf_Q R_1(P,D),
% \;\;\;\mbox{as }n\to\infty.
% $$
% Assuming that the infima above are achieved by
% probability distributions $\{Q^*_n\}$ and $Q^*$, 
% it is then natural to ask whether the (random)
% distribution $Q_n$ also converges $Q^*$ in some
% sense. In other words, when we are not only interested
% in the best compression rate $R_1(P,D)$ but also in
% how this rate con be achieved, it is natural to ask
% whether the distributions $Q_n^*$ correspond to 
% codebooks with near-optimal performance.
% 
% More generally, in the parametric estimation
% problem, suppose $\theta_n$ achieves the infimum
% in (\ref{e:R}), i.e.,
% $R_1^\Theta(P_{x_1^n},D)=
% R_1(P_{x_1^n},Q_{\theta_n},D)$,
% and that $\theta^*$ achieves the
% optimal limiting rate,
% $R_1^\Theta(P,D)=
% R_1(P,Q_{\theta^*},D)$. Our results
% so far give conditions under which 

Our proofs are in part based on a recent result
from \cite{Har:Gen:Sub}\cite{Har:Generalized:arXiv}.

\vskip 1ex
\begin{thm}
\cite{Har:Gen:Sub}\cite{Har:Generalized:arXiv}
\label{t:genAEP} If the source $(X_n)$ is stationary and 
ergodic with $X_1\sim P$, then 
\[ \liminf_{n\to\infty} R_1(P_{X_1^n},Q,D) \relas{=} R_1(P,Q,D) \]
for all $D\geq 0$ and
\begin{equation} \lim_{n\to\infty} R_1(P_{X_1^n},Q,D) 
\relas{=} R_1(P,Q,D) \label{e:genAEP} \end{equation}
for all $D$ in the set
\[ \Dc(P,Q) := \left\{ \! D\geq 0: R_1(P,Q,D) 
= \lim_{\lambda\uparrow 1} R_1(P,Q,\lambda D) \right\} . \]  
\end{thm}
\vskip 1ex

Similar to $\Dc(P)$, $\Dc(P,Q)$ always contains $0$ and any 
point where $R_1(P,Q,D)=\infty$.
Since the function $R_1(P,Q,D)$ is convex and nonincreasing in $D$ 
\cite{Har:Gen:Sub}\cite{Har:Generalized:arXiv}, 
$\Dc(P,Q)$ is the entire interval $[0,\infty)$, except perhaps
the single point where $R_1(P,Q,D)$ transitions from 
finite to infinite.

Somewhat loosely speaking,
the main point of this paper is to give conditions under which an 
infimum over $Q$ can be moved inside the limit 
in the above theorem.  
It turns out that our method of proof works equally well for moving 
an arg-infimum inside the limit.  The next theorem,
proved in Section~\ref{s:proofs},
is a strong consistency result giving conditions under
%which the approximate minimizers $(\theta_n)$ converge 
%to the optimal parameter $\theta^*$ corresponding to the 
%optimal reproduction distribution $Q_{\theta^*}$.
which the approximate minimizers $(\theta_n)$ converge 
to the optimal parameters $\{\theta^*\}$ corresponding to the 
optimal reproduction distributions $\{Q_{\theta^*}\}$.

\vskip 1ex
\begin{thm} \label{t:arginf}
Suppose the source $(X_n)$ is stationary and ergodic
with $X_1\sim P$,
the parameter set $\Theta$ is separable, 
and A1 and A2 hold. Then for all 
$D\in \Dc^\Theta(P)$, 
the set
\[ \Theta^* := \arginf_{\theta\in\Theta} R_1^\Theta(P,Q_\theta,D) \]
is not empty and any (typically random) sequence $(\theta_n)$ 
of approximate minimizers, i.e., satisfying,
\begin{equation} 
\limsup_{n\to\infty} 
R_1(P_{X_1^n},Q_{\theta_n},D) 
\leq \limsup_{n\to\infty} R_1^\Theta(P_{X_1^n},D),
\label{e:arginf} 
\end{equation}
has all of its limit points in $\Theta^*$ with probability 1.  
Furthermore, if $R_1^\Theta(P,D)<\infty$ and
either P2 or N2 holds,
then any sequence of approximate minimizers $(\theta_n)$ 
is relatively compact 
with probability 1.  Hence, $\theta_n\to\Theta^*$ with probability 1.
\end{thm}
\vskip 1ex

%%%%%%%%%%%%%%%%%%%%%%%%%%%%%%%%%%%%%%%%%%%%%%%%%
%%%%%%%%%%%%%%%%%%%%%%%%%%%%%%%%%%%%%%%%%%%%%%%%%
%%
% %  REMARKS AND EXTENSIONS:   RELATED ESTIMATORS
%%
%%%%%%%%%%%%%%%%%%%%%%%%%%%%%%%%%%%%%%%%%%%%%%%%%
%%%%%%%%%%%%%%%%%%%%%%%%%%%%%%%%%%%%%%%%%%%%%%%%%

\subsection{More General Estimators}
\label{s:general}

The upper and lower bounds of Theorems~\ref{t:ub} and~\ref{t:lb}
can be combined to extend our results to a variety of estimators
besides the ones considered already.   For example, instead of the
simple plug-in estimator, 
$$R_1^\Theta(P_{x_1^n},D)
=\inf_{\theta\in\Theta}R_1(P_{x_1^n},Q_\theta,D)$$
we may wish to consider
MDL-style penalized estimators, of the form,
\be
\inf_{\theta\in\Theta}\Big\{R_1(P_{x_1^n},Q_\theta,D)
+F_n(\theta)\Big\},
\label{eq:penalty}
\ee
for appropriate (nonnegative) penalty functions $F_n(\theta)$.
The penalty functions express our preference for
certain (typically less complex) subsets of $\Theta$
over others. This issue is, of course, particularly
important when estimating the optimal reproduction 
distribution as discussed in the previous section.
%i.e., when we are interested in the $\arginf$ in \eqref{eq:penalty}.\redend}
Note that in the case when no
distortion is allowed, these estimators reduce
to the classical ones used in lossless data compression
and in MDL-based model selection \cite{csiszar-shields:04}.
Indeed, if $A=\hat{A}$ are discrete sets, $\rho$ is Hamming
distance and $D=0$, then the estimator in (\ref{eq:penalty}) 
becomes,
$$
-\frac{1}{n}\;\sup_{\theta\in\Theta}\Big\{
\log Q_\theta^n(x_1^n)
-nF_n(\theta)\Big\},
$$
which is precisely the general form of a penalized 
maximum likelihood estimator.
% \redstart{(except with $\argsup$, of course).}\redend 
[As usual, $Q^n$ denotes 
the $n$-fold product
distribution on $\hat{A}^n$ corresponding to the 
marginal distribution $Q$.]

More generally, suppose we have a sequence of functions 
$(\varphi_n(x_1^n,\theta,D))$ with the properties 
that,
\begin{subequations}
\begin{gather}
\varphi_n(x_1^n,\theta,D) \geq R_1(P_{x_1^n},Q_\theta,D) \label{e:phi1} \\
\limsup_{n\to\infty} \varphi_n(X_1^n,\theta,D) \relas{=} 
\limsup_{n\to\infty} R_1(P_{X_1^n},Q_\theta,D) \label{e:phi2}
\end{gather}
\end{subequations}
for all $n$, $x_1^n$, $\theta$ and $D$.  
For each such sequence of functions $(\varphi_n)$,
we define a new estimator
for $R_1^\Theta(P,D)$ by,
\[ \varphi^\Theta_n(x_1^n,D) 
:= \inf_{\theta\in\Theta} \varphi_n(x_1^n,\theta,D) . \]
Condition \eqref{e:phi1} implies that any lower bound
for the plug-in estimator also holds here.
Also, by considering a single $\theta'$ for which 
$$\limsup_n R_1(P_{X_1^n},Q_{\theta'},D) 
\relas{\leq} R_1^\Theta(P,D)+\epsilon,$$ 
we see that \eqref{e:phi2} similarly implies
a corresponding upper bound. We thus obtain:

\vskip 1ex
\begin{cor} \label{c:phi} 
Theorems~\ref{t:ub}, \ref{t:lb} and~\ref{t:arginf} 
remain valid if $R_1^\Theta(P_{X_1^n},D)$ is replaced by 
$\varphi_n^\Theta(X_1^n,D)$ for any sequence of functions 
$(\varphi_n)$ satisfying \eqref{e:phi1} and \eqref{e:phi2}.
\end{cor}
\vskip 1ex

For example, the penalized plug-in estimators above
satisfy the conditions of the corollary, as long as 
the penalty functions $F_n$ satisfy, for each $\theta$,
$F_n(\theta)\to 0$ as $n\to\infty$.

Another example is the sequence of estimators 
based on the ``lossy likelihoods'' of
\cite{HarKon:Maximum:2002}, namely,
\[ \varphi_n(x_1^n,\theta,D) = 
-\frac{1}{n}\log Q_\theta^n\left(B_n(x_1^n,D)\right) \]
where $B_n(x_1^n,D)$ denotes the 
distortion-ball of radius $D$ centered at $x_1^n$,
\[ 
B_n(x_1^n,D) := 
\left\{y_1^n\in\hat{A}^n : \frac{1}{n}\sum_{k=1}^n \rho(x_k,y_k) \leq D\right\},
\]
cf.\ \cite{dembo-kontoyiannis}. 
Again, both conditions
\eqref{e:phi1} and \eqref{e:phi2} 
are valid in this case \cite{Har:Gen:Sub}\cite{Har:Generalized:arXiv}.  

%%%%%%%%%%%%%%%%%%%%%%%%%%%%%%%%%%%%%%%%%%%%%%%%%
%%%%%%%%%%%%%%%%%%%%%%%%%%%%%%%%%%%%%%%%%%%%%%%%%
%%
% %  REMARKS AND EXTENSIONS:   LLN
%%
%%%%%%%%%%%%%%%%%%%%%%%%%%%%%%%%%%%%%%%%%%%%%%%%%
%%%%%%%%%%%%%%%%%%%%%%%%%%%%%%%%%%%%%%%%%%%%%%%%%

\subsection{Nonstationary Sources}  \label{s:LLN}

As mentioned in the introduction, part of our motivation
comes from considering the problem of estimating the
rate-distortion function of distributions $P$ which 
cannot be computed analytically,
but which can be easily simulated by MCMC algorithms,
as is very often the case in image processing, for
example. Of course, MCMC samples are typically not stationary.
However, the distribution of the entire sequence of 
MCMC samples is dominated by (i.e., is absolutely continuous 
with respect to) a stationary and ergodic distribution, 
namely, the distribution of the same Markov chain started 
from its stationary distribution, which is of course the
target distribution  $P$.
Therefore, all of our results remain valid: Results that hold with 
probability 1 in the stationary case necessarily hold with 
probability 1 in the nonstationary case.  The only minor
technicality is that the initial distribution of the MCMC 
chain needs to be absolutely continuous with respect to $P$.

More generally (for non-Markov sources),
the requirements of stationarity and
ergodicity are more restrictive than necessary.
An inspection of the proofs (both here and in the 
proof of Theorem~\ref{t:genAEP} in \cite{Har:Gen:Sub}\cite{Har:Generalized:arXiv}),
reveals that we only need the source to have the
following law-of-large-numbers property:  

\begin{quote}
\item[LLN.]  There exists a random variable 
$X$ taking values in the source alphabet $A$,
such that,
\[ \textstyle  \frac{1}{n}\sum_{k=1}^n f(X_k) \relas{\to} E[f(X)], \]
for every nonnegative measurable function $f$.
\end{quote}

\vskip 1ex
\begin{thm} \label{t:ns} 
Theorems~\ref{t:ub}, \ref{t:lb} and~\ref{t:arginf}, Corollary \ref{c:phi} 
and the alternative conditions for A2 remain valid if, instead of being 
stationary and ergodic with $X_1\sim P$, the source merely satisfies the 
LLN property for some random variable $X\sim P$.  If the distortion measure 
$\rho$ is bounded, then the LLN property need only hold for bounded, 
measurable $f$.
\end{thm}
\vskip 1ex

Every stationary and ergodic source satisfies this LLN property as does any 
source whose distribution is dominated by the distribution of a stationary 
and ergodic source.  This LLN property is somewhat different from the 
requirement that the source be asymptotically mean stationary 
(a.m.s.)~with an ergodic mean stationary distribution 
\cite{GraKie:Asymptotically:1980}.  The latter is a stronger 
assumption in the sense that $f$ can depend on the entire future 
of the process, i.e., $n^{-1}\sum_{k=1}^n f(X_k,X_{k+1},\ldots) 
\relas{\to} E[f(X^\infty)]$, where 
$X^\infty$ is now a random variable on the infinite sequence space.  
It is a weaker assumption in that this convergence need only hold for 
bounded $f$.   The final statement of Theorem~\ref{t:ns} implies that our 
consistency results hold for a.m.s.~sources (with ergodic mean stationary 
distributions) as long as the distortion measure is bounded.

%%%%%%%%%%%%%%%%%%%%%%%%%%%%%%%%%%%%%%%%%%%%%%%%%
%%%%%%%%%%%%%%%%%%%%%%%%%%%%%%%%%%%%%%%%%%%%%%%%%
%%
% %  PROOFS
%%
%%%%%%%%%%%%%%%%%%%%%%%%%%%%%%%%%%%%%%%%%%%%%%%%%
%%%%%%%%%%%%%%%%%%%%%%%%%%%%%%%%%%%%%%%%%%%%%%%%%
 
\section{Proofs} \label{s:proofs}
 
We frequently use the alternative 
representation \cite{Har:Gen:Sub}\cite{Har:Generalized:arXiv}
\begin{equation} 
R_1(P,Q,D) = 
\sup_{\lambda\leq 0}\Bigl[\lambda D - 
E_{X\sim P}\bigl[\log E_{Y\sim Q} e^{\lambda\rho(X,Y)}\bigr]\Bigr] 
\label{e:RPQD} 
\end{equation} 
which is valid for all choices of $P$, $Q$ and $D$.  

This representation makes it easy to prove that
\begin{equation} R_1^\Theta(\epsilon P'+(1-\epsilon)P,D) \geq \epsilon R_1^\Theta(P',D/\epsilon) \label{e:Reps} \end{equation}
for $\epsilon\in(0,1)$, which is used above in Example~\ref{ex:dc}.
Indeed, 
\begin{align*}
& R_1(\epsilon P' + (1-\epsilon)P,Q_\theta,D) \\
& \quad = \sup_{\lambda \leq 0} \Bigl[\lambda D - \epsilon 
E_{X \sim P'}\bigl[\log E_{Y\sim Q_\theta} e^{\lambda\rho(X,Y)}\bigr]  %\\
%& \quad \quad  
- (1-\epsilon) E_{X\sim P}\bigl[\log E_{Y\sim Q_\theta} e^{\lambda\rho(X,Y)}\bigr]\Bigr] \\
& \quad \geq \sup_{\lambda \leq 0} \Bigl[\lambda D - \epsilon 
E_{X\sim P'}\bigl[\log E_{Y\sim Q_\theta} [e^{\lambda\rho(X,Y)}]\bigr] \Bigr] \\
& \quad = \epsilon \sup_{\lambda \leq 0} \Bigl[\lambda D/\epsilon - 
E_{X\sim P'}\bigl[\log E_{Y\sim Q_\theta} [e^{\lambda\rho(X,Y)}]\bigr]\Bigr] \\
& \quad = \epsilon R_1(P',Q_\theta,D/\epsilon). \end{align*}
Taking the infimum over $\theta\in\Theta$ on both sides gives \eqref{e:Reps}.

%%%%%%%%%%%%%%%%%%%%%%%%%%%%%%%%%%%%%%%%%%%%%%%%%
%%%%%%%%%%%%%%%%%%%%%%%%%%%%%%%%%%%%%%%%%%%%%%%%%
%%
% %  PROOFS:   MEASURABILITY
%%
%%%%%%%%%%%%%%%%%%%%%%%%%%%%%%%%%%%%%%%%%%%%%%%%%
%%%%%%%%%%%%%%%%%%%%%%%%%%%%%%%%%%%%%%%%%%%%%%%%%

\subsection{Measurability} \label{s:meas}

Here we discuss the various measurability assumptions that are used 
throughout the paper.  Note that we do not always establish the 
measurability of an event if it contains another measurable event 
that has probability 1.

Since $\rho$ is product measurable, $x\mapsto 
E_\theta [e^{\lambda\rho(x,Y)}]$ is measurable.  This implies that 
$x_1^n \mapsto \lambda D - E_{P_{x_1^n}} 
\Big\{\log E_\theta [e^{\lambda\rho(X,Y)}]\Big\}$ 
is measurable.  Since this is concave in $\lambda$ \cite{Har:Generalized:arXiv}, we can 
evaluate the supremum over all $\lambda\leq 0$ 
in (\ref{e:RPQD}) by considering only countably many 
$\lambda\leq 0$, which means that $x_1^n 
\mapsto R_1(P_{x_1^n},Q_\theta,D)$ is measurable.   

If $\Theta$ is a separable metric space and $f:\Theta\times A^n \to \bar{\mathbb{R}}$ is measurable for fixed $\theta\in\Theta$ and continuous for fixed $x_1^n\in A^n$, then $x_1^n\mapsto\sup_{\theta\in U} f(\theta,x_1^n)$ is measurable for any subset $U\subseteq\Theta$.  This is because $\sup_{\theta\in U} f = \sup_{\theta\in U'} f$ for any (at most) countable dense subset $U'\subseteq U$, and the latter is measurable because $U'$ is (at most) countable.  Since $\Theta$ is separable, such a $U'$ always exists, and since $f(\cdot,x_1^n)$ is continuous, $U'$ can be chosen independently of $x_1^n$.  An identical argument holds for $\inf_{\theta\in U} f$.  We make use of this frequently in the lower bound, where the necessary continuity comes from A1.

%%%%%%%%%%%%%%%%%%%%%%%%%%%%%%%%%%%%%%%%%%%%%%%%%
%%%%%%%%%%%%%%%%%%%%%%%%%%%%%%%%%%%%%%%%%%%%%%%%%
%%
% %  PROOFS:   UPPER BOUND
%%
%%%%%%%%%%%%%%%%%%%%%%%%%%%%%%%%%%%%%%%%%%%%%%%%%
%%%%%%%%%%%%%%%%%%%%%%%%%%%%%%%%%%%%%%%%%%%%%%%%%

\subsection{Proof of Theorem~\ref{t:ub}}
The upper bound in Theorem~\ref{t:ub} 
is deduced from Theorem~\ref{t:genAEP} as follows.
If $D=0$ or $R_1^\Theta(P,D)=\infty$, then choose 
$D'=D$, otherwise, choose $D' < D$ such that 
$R_1^\Theta(P,D')\leq R_1^\Theta(P,D)+\epsilon/2$.  We can always do this 
since $D\in\Dc^\Theta(P)$.  Now pick $\theta\in\Theta$ with 
$R_1(P,Q_\theta,D') \leq R_1^\Theta(P,D')+\epsilon/2$.  This ensures that 
$D\in\Dc(P,Q_\theta)$ and Theorem~\ref{t:genAEP} gives
\begin{align*} 
& \limsup_{n\to\infty} R_1^\Theta(P_{X_1^n},D) \leq 
\limsup_{n\to\infty} R_1(P_{X_1^n},Q_\theta,D) %\notag \\ 
%& \quad 
\relas{=} R_1(P,Q_\theta,D) \leq R_1(P,Q_\theta,D') 
\leq R_1^\Theta(P,D)+\epsilon 
\end{align*}
completing the proof.  Notice that if we switch the $\limsup$ to a $\liminf$, 
we can remove any restrictions on $D$ since there are no restrictions 
in this case in Theorem~\ref{t:genAEP}.  This gives \eqref{e:liminf}.

%%%%%%%%%%%%%%%%%%%%%%%%%%%%%%%%%%%%%%%%%%%%%%%%%
%%%%%%%%%%%%%%%%%%%%%%%%%%%%%%%%%%%%%%%%%%%%%%%%%
%%
% %  PROOFS:   LOWER BOUND
%%
%%%%%%%%%%%%%%%%%%%%%%%%%%%%%%%%%%%%%%%%%%%%%%%%%
%%%%%%%%%%%%%%%%%%%%%%%%%%%%%%%%%%%%%%%%%%%%%%%%%

\subsection{Proof of Theorem~\ref{t:lb}}
\label{s:lb}

Here we prove the lower bound of Theorem~\ref{t:lb}.  
Let $\tau$ denote the metric on $\Theta$ and let 
$O(\theta,\epsilon):=\{\theta':\tau(\theta',\theta) < \epsilon\}$ denote 
the open ball of radius $\epsilon$ centered at $\theta$.
The main goal is to prove that
\begin{equation} 
\lim_{\epsilon\downarrow 0} \liminf_{n\to\infty} 
\inf_{\theta'\in O(\theta,\epsilon)} 
R_1(P_{X_1^n},Q_{\theta'},D) 
\relas{\geq} R_1(P,Q_\theta,D) 
\label{e:epi} 
\end{equation}
for all $\theta\in\Theta$ simultaneously, that is, the exceptional set 
can be chosen independently of $\theta$.  To see how this gives the lower 
bound, first choose a sequence $(\theta_n)$ according to A2 and 
a subsequence $(n_k)$ along which the $\liminf$ on the left side 
of \eqref{e:A2} is actually a limit.  Let $\theta^*$ be a limit point 
of the subsequence $(\theta_{n_k})$.  Note that such a 
$\theta^*$ exists with probability 1 by assumption A2 
and that it depends on $X_1^\infty$.  We have,
\begin{align} 
& \liminf_{n\to\infty} R_1^\Theta(P_{X_1^n},D) 
\geq \liminf_{n\to\infty} R_1(P_{X_1^n},Q_{\theta_n},D) 
= \lim_{k\to\infty} R_1(P_{X_1^{n_k}},Q_{\theta_{n_k}},D) 
\notag \\ & \quad \relas{\geq} \liminf_{n\to\infty} 
\inf_{\theta'\in O(\theta^*,\epsilon)} 
R_1(P_{X_1^n},Q_{\theta'},D) 
\label{e:epi1} 
\end{align}
for each $\epsilon > 0$.   The first inequality is from \eqref{e:A2} 
and the last is valid because 
infinitely many elements of $(\theta_{n_k})$ are in 
$O(\theta^*,\epsilon)$ for any $\epsilon > 0$.  
Letting $\epsilon\downarrow 0$ in \eqref{e:epi1} and 
using \eqref{e:epi} gives
\[ 
\liminf_{n\to\infty} R_1^\Theta(P_{X_1^n},D) 
\relas{\geq} R_1(P,Q_{\theta^*},D) \geq R_1^\Theta(P,D) 
\]
as desired.  Note that with \eqref{e:liminf} this also implies that that $\theta^*$ 
achieves the infimum in the definition of $R_1^\Theta(P,D)$.

We need only prove \eqref{e:epi}.  For any $\lambda \leq 0$, $\theta\in\Theta$ and $\epsilon > 0$, the pointwise ergodic theorem gives 
\begin{align} & \lim_{n\to\infty} \frac{1}{n}\sum_{k=1}^n \sup_{\theta'\in O(\theta,\epsilon)} \log E_{\theta'}[e^{\lambda\rho(X_k,Y)}] %\notag \\ & \quad 
\relas{=} E_P\left[\sup_{\theta'\in O(\theta,\epsilon)} \log 
E_{\theta'} [e^{\lambda\rho(X,Y)}]\right] . \label{e:epi2} \end{align}
(See Section~\ref{s:meas} for measurability.)
Fix an at most countable, dense subset $\tilde\Theta\subseteq\Theta$.  We can choose the exceptional sets in \eqref{e:epi2} independently of $\theta\in\tilde\Theta$ and $\epsilon>0$ rational.  For any $\theta\in\Theta$ and $\epsilon > 0$ we can choose a $\tilde\theta\in\tilde\Theta$ and a rational $\tilde\epsilon>\epsilon$ such that $O(\theta,\epsilon)\subseteq O(\tilde\theta,\tilde\epsilon)\subseteq O(\theta,2\epsilon)$.  Since the exceptional sets in \eqref{e:epi2} do not depend on $\tilde\theta$ and $\tilde\epsilon$, we have that 
\begin{align} & \limsup_{n\to\infty} \sup_{\theta'\in O(\theta,\epsilon)} \frac{1}{n}\sum_{k=1}^n \log E_{\theta'}[e^{\lambda\rho(X_k,Y)}] %\notag \\ & \quad 
\leq
 \limsup_{n\to\infty} \frac{1}{n}\sum_{k=1}^n \sup_{\theta'\in O(\theta,\epsilon)} \log E_{\theta'}[e^{\lambda\rho(X_k,Y)}] \notag \\ & \quad \leq \limsup_{n\to\infty} \frac{1}{n}\sum_{k=1}^n \sup_{\theta'\in O(\tilde\theta,\tilde\epsilon)} \log E_{\theta'}[e^{\lambda\rho(X_k,Y)} ]\notag %\\ & \quad 
 \relas{=} 
E_P\left[\sup_{\theta'\in O(\tilde\theta,\tilde\epsilon)} \log 
E_{\theta'} [e^{\lambda\rho(X,Y)}]\right] \notag \\ 
& \quad \leq E_P\left[\sup_{\theta'\in O(\theta,2\epsilon)} \log 
E_{\theta'} [e^{\lambda\rho(X,Y)}]\right] \label{e:epi3} \end{align}
simultaneously for all $\theta\in\Theta$ and $\epsilon > 0$, that is, the exceptional set can be chosen independently of $\theta$ and $\epsilon$. 

The monotone convergence theorem and the continuity in A1 give 
\begin{align*} & \lim_{\epsilon\downarrow 0} E_P\left[\sup_{\theta'\in O(\theta,2\epsilon)} \log E_{\theta'} [e^{\lambda\rho(X,Y)}]\right] %\\ & \quad 
= 
E_P\left[\lim_{\epsilon\downarrow 0} \sup_{\theta'\in O(\theta,2\epsilon)} \log E_{\theta'} [e^{\lambda\rho(X,Y)}]\right]  %\\ & \quad 
= 
E_P\left[\log E_\theta [e^{\lambda\rho(X,Y)}]\right] . \end{align*}
Combining this with \eqref{e:epi3} and letting $\epsilon\downarrow 0$ gives
\begin{align} & \lim_{\epsilon\downarrow 0} \limsup_{n\to\infty} \sup_{\theta'\in O(\theta,\epsilon)} \frac{1}{n} \sum_{k=1}^{n} \log 
E_{\theta'} [e^{\lambda\rho(X_k,Y)}] %\notag \\ & \quad 
\relas{\leq} 
E_P\left[\log E_\theta [e^{\lambda\rho(X,Y)}]\right] \label{e:epi4} \end{align}
simultaneously for all $\theta\in\Theta$.  

Both sides of \eqref{e:epi4} are nondecreasing with $\lambda$.
Furthermore, the right side of \eqref{e:epi4} is continuous from above 
for $\lambda < 0$.  (To see this, use the dominated convergence theorem 
to move the limit through $E_\theta$ and the monotone convergence theorem 
to move the limit through $E_P$.)  These two facts imply
that we can also choose the exceptional sets independently of $\lambda\leq 0$ (by first applying \eqref{e:epi4} for $\lambda$ rational and then squeezing).  Applying \eqref{e:epi4} to the representation in \eqref{e:RPQD} gives, 
for each $\lambda \leq 0$,
\begin{align*} & \lim_{\epsilon\downarrow 0} \liminf_{n\to\infty} \inf_{\theta'\in O(\theta,\epsilon)} R_1(P_{X_1^n},Q_{\theta'},D)  
%\notag \\ & \quad 
\geq \lim_{\epsilon\downarrow 0} \liminf_{n\to\infty} \inf_{\theta'\in O(\theta,\epsilon)} \left[\lambda D - \frac{1}{n}\sum_{k=1}^n \log 
E_{\theta'} [e^{\lambda\rho(X_k,Y)}] \right] 
\notag \\ & \quad = \lambda D - \lim_{\epsilon\downarrow 0} \limsup_{n\to\infty} \sup_{\theta'\in O(\theta,\epsilon)} \frac{1}{n}\sum_{k=1}^n \log 
E_{\theta'} [e^{\lambda\rho(X_k,Y)}] %\notag \\ & \quad 
\relas{\geq} \lambda D - E_P\left[\log E_\theta [e^{\lambda\rho(X,Y)}]\right]  \end{align*}
simultaneously for all $\theta\in\Theta$ and $\lambda\leq 0$.  
Optimizing over $\lambda\leq 0$ on the right gives \eqref{e:epi}.

%%%%%%%%%%%%%%%%%%%%%%%%%%%%%%%%%%%%%%%%%%%%%%%%%
%%%%%%%%%%%%%%%%%%%%%%%%%%%%%%%%%%%%%%%%%%%%%%%%%
%%
% %  PROOFS:   ALTERNATIVE ASSUMPTIONS
%%
%%%%%%%%%%%%%%%%%%%%%%%%%%%%%%%%%%%%%%%%%%%%%%%%%
%%%%%%%%%%%%%%%%%%%%%%%%%%%%%%%%%%%%%%%%%%%%%%%%%

\subsection{Alternative Assumptions}

Here we discuss the various alternative assumptions that imply A1 and A2.  P1 implies A1 because $y\mapsto e^{\lambda\rho(x,y)}$ is bounded and measurable for each $x\in A$ and $\lambda\leq 0$.  N1 implies A1 because $y\mapsto e^{\lambda\rho(x,y)}$ is bounded and continuous for each $x\in A$ and $\lambda\leq 0$.   

%%%%%%%%%%%%%%%%%%%%%%%%%%%%%%%%%%%%%%%%%%%%%%%%%
%%%%%%%%%%%%%%%%%%%%%%%%%%%%%%%%%%%%%%%%%%%%%%%%%
%%
% %  PROOFS:   ALTERNATIVE ASSUMPTIONS:   P2 IMPLIES A2
%%
%%%%%%%%%%%%%%%%%%%%%%%%%%%%%%%%%%%%%%%%%%%%%%%%%
%%%%%%%%%%%%%%%%%%%%%%%%%%%%%%%%%%%%%%%%%%%%%%%%%

\subsubsection{P2 Implies A2}

Here we prove that P2 implies A2 when $(X_n)$ is stationary and ergodic 
with $X_1\sim P$.  Fix $D$, $\Delta$ and $K$ according to P2, so that 
$T_\epsilon := \{\theta : Q_\theta(B(K,D+\Delta)) \geq \epsilon\}$ 
is relatively compact for each $\epsilon > 0$.  
We will first show that
\begin{equation} \lim_{\epsilon\downarrow 0} \liminf_{n\to\infty} 
\inf_{\theta\in T_\epsilon^c} R_1(P_{X_1^n},Q_\theta,D) \relas{=} 
\infty \label{e:P2A2:1a}  \end{equation}
where $T_\epsilon^c$ denotes the complement of $T_\epsilon$.  If $T_\epsilon^c$ is empty for some $\epsilon > 0$, then \eqref{e:P2A2:1a} follows from the convention that $\inf\emptyset=\infty$.  We can thus focus on the case where $T_\epsilon^c$ is not empty for all $\epsilon > 0$.

Define $\lambda_\epsilon := (\log\epsilon)/(D+\Delta)$.  Since 
\[ \rho(x,y) \geq (D+\Delta)\ind\{x\in K, y\in B(K,D+\Delta)^c\} \]
we have for any $\theta\in T_\epsilon^c$ 
\begin{align*} & \log E_\theta [e^{\lambda_\epsilon \rho(x,Y)}] 
\leq \ind\{x\in K\}\log\left[\epsilon+e^{\lambda_\epsilon(D+\Delta)}\right] %\\ & \quad 
= \ind\{x\in K\}\log(2\epsilon) . \end{align*}
This and the representation in \eqref{e:RPQD} imply that
\begin{align*} & \inf_{\theta\in T_\epsilon^c} R_1(P_{X_1^n},Q_\theta,D) 
%\\ & \quad 
\geq \inf_{\theta\in T_\epsilon^c} \left[ \lambda_\epsilon D - \frac{1}{n}\sum_{k=1}^n \log E_\theta [e^{\lambda_\epsilon \rho(X_k,Y)}] \right]
\\ & \quad 
\geq \frac{D}{D+\Delta}\log \epsilon - \frac{1}{n}\sum_{k=1}^n  \ind\{X_k\in K\}\log(2\epsilon) .
\end{align*}
Taking limits, the pointwise ergodic theorem gives
\begin{align} & \liminf_{n\to\infty} \inf_{\theta\in T_\epsilon^c} R_1(P_{X_1^n},Q_\theta,D) 
%\notag \\ & \quad 
\relas{\geq} \frac{D}{D+\Delta}\log \epsilon - P(K)\log(2\epsilon)  . \label{e:P2A2:erg} \end{align}
Letting $\epsilon \downarrow 0$ ($\epsilon$ rational) and noting that $P(K) > D/(D+\Delta)$ by assumption gives \eqref{e:P2A2:1a}.

Now we will show that \eqref{e:P2A2:1a} implies A2.   
Fix a realization $x_1^\infty$ of $X_1^\infty$ for which 
\eqref{e:P2A2:1a} holds.  Let $(n_k)$ be a subsequence for which 
\[ L := \liminf_{n\to\infty} R_1^\Theta(P_{x_1^n},D) = \lim_{k\to\infty} R_1^\Theta(P_{x_1^{n_k}},D). \]
If $L=\infty$, we can simply take $\theta_n=\theta$ for any constant $\theta$ and all $n$.  If $L < \infty$, choose $\theta_{n_k}$ so that 
\[ \lim_{k\to\infty} R_1(P_{x_1^{n_k}},Q_{\theta_{n_k}},D) = L . \] 
Then \eqref{e:P2A2:1a} implies that there exists an $\epsilon > 0$ for 
which $\theta_{n_k}$ must be in $T_\epsilon$ for all $k$ large enough.  
Since $T_\epsilon$ has compact closure, the subsequence 
$(\theta_{n_k})$ is relatively compact and it can always be 
embedded in a relatively compact sequence $(\theta_n)$.   
Since $x_1^\infty$ is (with probability 1) arbitrary, 
the proof is complete.

%%%%%%%%%%%%%%%%%%%%%%%%%%%%%%%%%%%%%%%%%%%%%%%%%
%%%%%%%%%%%%%%%%%%%%%%%%%%%%%%%%%%%%%%%%%%%%%%%%%
%%
% %  PROOFS:   ALTERNATIVE ASSUMPTIONS:   N2 IMPLIES A2
%%
%%%%%%%%%%%%%%%%%%%%%%%%%%%%%%%%%%%%%%%%%%%%%%%%%
%%%%%%%%%%%%%%%%%%%%%%%%%%%%%%%%%%%%%%%%%%%%%%%%%

\subsubsection{N2 Implies A2}

Here we prove that N2 implies A2 when $(X_n)$ is stationary and ergodic 
with $X_1\sim P$.  For each $\epsilon > 0$ and each $M > 0$, let 
$K(\epsilon,M)$ be the set in N2.  The pointwise ergodic theorem gives,
\begin{equation} 
\lim_{n\to\infty} P_{X_1^n}(K(\epsilon,M)) \relas{=} P(K(\epsilon,M)).
\label{e:K} 
\end{equation}
Fix a realization $x_1^\infty$ of $X_1^\infty$ for which \eqref{e:K} holds 
for all rational $\epsilon$ and $M$.  Let $(n_k)$ be a subsequence for 
which 
\[ 
L := \liminf_{n\to\infty} R_1^\Theta(P_{x_1^n},D) 
= \lim_{k\to\infty} R_1^\Theta(P_{x_1^{n_k}},D). 
\]
If $L=\infty$, we can simply take $\theta_n=\theta$ 
for any constant $\theta$ and all $n$.  If $L < \infty$,  
for $k$ large enough both sides are 
finite and we can choose $W_k \in W(P_{x_1^{n_k}},D)$ so that 
\[ H(W_k\|W_k^A\times W_k^{\hat{A}}) \leq R_1^\Theta(P_{x_1^{n_k}},D) + 1/k . \]
Let $Q_{\theta_{n_k}} = W_k^{\hat{A}}$ and note that
\[ R_1(P_{x_1^{n_k}},Q_{\theta_{n_k}},D) 
\leq R_1^\Theta(P_{x_1^{n_k}},D)+ 1/k  . \] 
We will show that $\theta_{n_k}$ is relatively compact by showing that 
the sequence $(Q_k := Q_{\theta_{n_k}})$ is tight.\footnote{A sequence 
	of probability measures $(Q_k)$ on 
	$(\hat{A},\mathcal{\hat{A}})$ is said to be {\em tight} if 
	$\sup_F \liminf_{k\to\infty} Q_k(F) = 1$,
	where the supremum is over all compact (measurable) 
	$F\subseteq \hat{A}$.
	If $(Q_k)$ is tight, then Prohorov's Theorem states 
	that it is relatively compact in the topology 
	of weak convergence of probability 
	measures \cite{Kal:Foundations:2002}.}
This will complete the proof just like in the previous section.

Fix $\epsilon > 0$ rational and $M > 2D/\epsilon$ rational.  Let $K=K(\epsilon/2,M)$.  We have
\begin{align*} &  D \geq E_{(U,V)\sim W_k}[\rho(U,V)] \geq MW_k(K\times B(K,M)^c) %\\ & \quad 
\geq 2DW_k(K\times B(K,M)^c)/\epsilon . \end{align*}
This implies that $W_k(K\times B(K,M)^c) \leq \epsilon/2$ and we can bound
\begin{align*} & Q_k(B(K,M)) = W_k^{\hat{A}}(B(K,M)) \geq W_k(K\times B(K,M))  %\\ & \quad 
= P_{x_1^{n_k}}(K) - W_k(K\times B(K,M)^c) 
\\ & \quad
\geq P_{x_1^{n_k}}(K) - \epsilon/2 . \end{align*}
Taking limits and applying \eqref{e:K} gives
\[ \liminf_{n\to\infty} Q_k(B(K,M)) \geq P(K)-\epsilon/2 > 1-\epsilon . \]
Since $B(K,M)$ has compact closure and since $\epsilon$ was arbitrary, 
the sequence $(Q_k)$ is tight.

%%%%%%%%%%%%%%%%%%%%%%%%%%%%%%%%%%%%%%%%%%%%%%%%%
%%%%%%%%%%%%%%%%%%%%%%%%%%%%%%%%%%%%%%%%%%%%%%%%%
%%
% %  PROOFS:   CONVERGENCE OF MINIMIZERS
%%
%%%%%%%%%%%%%%%%%%%%%%%%%%%%%%%%%%%%%%%%%%%%%%%%%
%%%%%%%%%%%%%%%%%%%%%%%%%%%%%%%%%%%%%%%%%%%%%%%%%

\subsection{Proof of Theorem~\ref{t:arginf}}

Here we prove the convergence-of-minimizers 
result given in Theorem~\ref{t:arginf}.  The proof of Theorem \ref{t:lb} in Section~\ref{s:lb} shows that $\Theta^*$ is not empty.   
The assumptions ensure that both the lower 
and upper bounds for consistency of the
plug-in estimator hold, so that
$R_1^\Theta(P_{X_1^n},D)\relas{\to}R_1^\Theta(P,D)$.  This shows that 
any sequence $(\theta_n)$ satisfying \eqref{e:arginf} also satisfies 
\eqref{e:A2} with probability 1, and that the $\limsup$ and the 
$\liminf$ agree.  Let $\theta^*$ be any limit point of this sequence 
(if one exists).  Following the steps at the beginning of the proof 
of Theorem~\ref{t:lb} in Section~\ref{s:lb}, we see that $\theta^*\in\Theta^*$.  

Now further suppose that $R_1^\Theta(P,D)$ is finite so that
\begin{equation} R_1(P_{X_1^n},Q_{\theta_n},D) \relas{\to} R_1^\Theta(P,D) < M < \infty . \label{e:arginf:1} \end{equation}  We want to show that 
the sequence $(\theta_n)$ is relatively compact with probability 1.
If P2 holds, then \eqref{e:P2A2:1a} immediately implies that there exists an $\epsilon > 0$ such that $\theta_n\in T_\epsilon$ eventually, 
with probability 1.
Since $T_\epsilon$ is relatively compact, so is $(\theta_n)$.  

Alternatively, suppose N2 holds.  To show that $(\theta_n)$ is 
relatively compact with probability 1, we need only show that 
$(Q_{\theta_n})$ is tight w.p.1.  Fix a realization 
$x_1^\infty$ where the convergence in \eqref{e:arginf:1} holds, 
where \eqref{e:K} holds for all rational $\epsilon$ and $M$, and 
where $R_1^\Theta(P_{x_1^n},D) \to R_1^\Theta(P,D)$.  
For $n$ large enough, the left side of \eqref{e:arginf:1} is finite, 
so $W(P_{x_1^n},D)$ is not empty and we can choose a sequence 
$(W_n)$ with $W_n\in W(P_{x_1^n},D)$ so that
\[ H(W_n\|P_{x_1^n}\times Q_{\theta_n}) \to R_1^\Theta(P,D) . \]
Let $Q_n := W_n^{\hat{A}}$.  An inspection of the above proof that N2 
implies A2 shows that the sequence $(Q_n)$ is tight.     
We will show that 
$H(Q_n\|Q_{\theta_n})\to 0$, implying that $(Q_{\theta_n})$ is also tight 
(because, for example, relative entropy bounds total variation distance).  
Indeed, 
\begin{align*}  & \underbrace{H(W_n\|P_{x_1^n}\times Q_{\theta_n})}_{a_n} = H(W_n\|P_{x_1^n}\times W_n^{\hat{A}}) + H(W_n^{\hat{A}}\|Q_{\theta_n}) %\\ & \quad 
\geq 
\underbrace{R_1^\Theta(P_{x_1^n},D)}_{b_n} + \underbrace{H(Q_n\|Q_{\theta_n})}_{c_n} . \end{align*}
Since $a_n$ and $b_n$ both converge to $R_1^\Theta(P,D)$, which is finite, $c_n\to 0$, as claimed.

%%%%%%%%%%%%%%%%%%%%%%%%%%%%%%%%%%%%%%%%%%%%%%%%%
%%%%%%%%%%%%%%%%%%%%%%%%%%%%%%%%%%%%%%%%%%%%%%%%%
%%
% %  PROOFS:   LAW OF LARGE NUMBERS
%%
%%%%%%%%%%%%%%%%%%%%%%%%%%%%%%%%%%%%%%%%%%%%%%%%%
%%%%%%%%%%%%%%%%%%%%%%%%%%%%%%%%%%%%%%%%%%%%%%%%%

\subsection{Proof of Theorem~\ref{t:ns}}

Here we prove the result of Theorem~\ref{t:ns}, based on
the law-of-large-numbers property.
Inspecting all of the proofs in this paper 
reveals that the assumption of a stationary and ergodic source is only used 
to invoke the pointwise ergodic theorem.  Furthermore, the pointwise ergodic 
theorem is not needed in full generality, only the LLN property is used.  
The relevant equations are \eqref{e:epi2}, \eqref{e:P2A2:erg} 
and~\eqref{e:K}.  Note that if $\rho$ is bounded, then it is enough to 
have the LLN property hold for bounded $f$.

Equation~\eqref{e:genAEP} from Theorem~\ref{t:genAEP}, which we used in the proof of the upper bound, also assumes a stationary and ergodic source.  The proof of a more general result than Theorem~\ref{t:genAEP} is in \cite{Har:Gen:Sub}\cite{Har:Generalized:arXiv}, but that result makes extensive use of the stationarity assumption.  A careful reading reveals that only the LLN property is needed for \eqref{e:genAEP}.  For completeness, we will give a proof, referring only to \cite{Har:Generalized:arXiv} for results that do not depend on the nature of the source.   Specifically, what we need to prove for the upper bound is that
\begin{equation} 
\limsup_{n\to\infty} R_1(P_{X_1^n},Q,D) \relas{\leq} R_1(P,Q,D) \label{e:genAEP:LLN}
\end{equation}
for all $D\in\Dc(P,Q)$.

If the source satisfies the LLN property for a 
random variable $X$ with distribution $P$, then
\begin{align} & \lim_{n\to\infty} \frac{1}{n}\sum_{k=1}^n \log 
E_{Y\sim Q} [e^{\lambda\rho(X_k,Y)}]
%\notag \\ & \quad 
\relas{=} E_{X\sim P}\left[\log E_{Y\sim Q} 
[e^{\lambda\rho(X,Y)}]\right]  
:= \Lambda(\lambda) . \label{e:ns:lam} \end{align}
Furthermore, since both sides are monotone in $\lambda$, the exceptional sets can be chosen independently of $\lambda$.   The LLN property also implies that
\begin{align} & \lim_{n\to\infty} \frac{1}{n} \sum_{k=1}^n 
E_{Y\sim Q} [\rho(X_k,Y)]
%\notag \\ & \quad 
\relas{=} E_{X\sim P}\left[E_{Y\sim Q} [\rho(X,Y)]\right] 
:= \Dave  \label{e:ns:Dave} . \end{align}
Note that if $\rho$ is bounded, then the LLN property need only hold for bounded $f$ in both \eqref{e:ns:lam} and \eqref{e:ns:Dave}.  

Define $\Lambda^*(D) := \sup_{\lambda \leq 0} \left[\lambda D - \Lambda(\lambda)\right]$ and $\Dmin:=\inf\{D\geq 0: \Lambda^*(D) < \infty\}$,
with the convention that the infimum of the empty set equals
$+\infty$. In \cite{Har:Generalized:arXiv} it is shown that 
$\Dmin\leq\Dave$, that $\Lambda^*$ is convex, nonincreasing 
and continuous from the right, and that
\[ \Lambda^*(D) = \begin{cases} \infty  & \text{if $D < \Dmin$} \\ \text{strictly convex} & \text{if $\Dmin < D < \Dave$} \\ 0 & \text{if $D \geq \Dave$} \end{cases} \]
where some of these cases may be empty.  Notice that $\Lambda^*$ is continuous except perhaps at $\Dmin$, where it will not be continuous from the left if $\Lambda^*(\Dmin) < \infty$.

Fix a realization $x_1^\infty$ of $X_1^\infty$ for which \eqref{e:ns:lam} holds for all $\lambda$ and for which \eqref{e:ns:Dave} holds.  
Define the random variables 
\[ Z_n := \frac{1}{n}\sum_{k=1}^n \rho(x_k,Y_k) \]
for $n\geq 1$, where the sequence $(Y_k)$ consists of independent and 
identically distributed (i.i.d.) random variables with common distribution 
$Q$. Then \eqref{e:ns:lam} implies that
\begin{equation} \lim_{n\to\infty} \frac{1}{n}\log E[e^{\lambda n Z_n}] = \Lambda(\lambda) . \label{e:Zlim} \end{equation}

We will first show that
\begin{equation} \lim_{n\to\infty} -\frac{1}{n}\log \Prob\{Z_n\leq D\} = \Lambda^*(D) = R(P,Q,D) \label{e:Z} \end{equation}
for all $D\geq 0$ except the special case 
when both $D=\Dmin$ and $\Lambda^*(\Dmin)<\infty$.  
The second equality in \eqref{e:Z} is always valid 
\cite{Har:Gen:Sub}\cite{Har:Generalized:arXiv}.  If $D < \Dmin$, or $D=\Dmin$ and $\Lambda^*(\Dmin)=\infty$, or $\Dmin < D \leq \Dave$, the first equality in \eqref{e:Z} follows from 
\cite[Lemma 11]{Har:Generalized:arXiv}, which is a slight modification of the G\"artner-Ellis Theorem in the theory of large deviations.  The aforementioned properties of $\Lambda^*$ and the convergence in \eqref{e:Zlim} are what we need to use 
\cite[Lemma 11]{Har:Generalized:arXiv}.  If $D > \Dave$, then $\Lambda^*(D)=0$ and we need only show that $\liminf_n \Prob\{Z_n \leq D\} > 0$.  But this follows from 
Chebychev's inequality and \eqref{e:ns:Dave} because
\begin{align*} &  \Prob\{Z_n \leq D\} = 1-\Prob\{Z_n > D\} \geq 1 - E[Z_n]/D %\\ & \quad 
\to 1 - \Dave/D > 0 . \end{align*} 
This proves \eqref{e:Z}, except for the special case
when $D=\Dmin$ and $\Lambda^*(\Dmin)<\infty$ -- which
exactly corresponds to $D\not\in\Dc(P,Q)$.

Finally, \eqref{e:Z} gives \eqref{e:genAEP:LLN} because \cite{Har:Gen:Sub}\cite{Har:Generalized:arXiv}
\[ R_1(P_{x_1^n},Q,D)  \leq -\frac{1}{n}\log \Prob\{Z_n\leq D\} \]
and because $x_1^\infty$ is (with probability 1) arbitrary. 

%%%%%%%%%%%%%%%%%%%%%%%%%%%%%%%%%%%%%%%%%%%%%%%%%
%%%%%%%%%%%%%%%%%%%%%%%%%%%%%%%%%%%%%%%%%%%%%%%%%
%%
% %  ACKNOWLEDGEMENTS AND BIBLIOGRAPHY
%%
%%%%%%%%%%%%%%%%%%%%%%%%%%%%%%%%%%%%%%%%%%%%%%%%%
%%%%%%%%%%%%%%%%%%%%%%%%%%%%%%%%%%%%%%%%%%%%%%%%%

\section*{Acknowledgments}
The authors wish to thank M.~Madiman for many useful comments.

\bibliographystyle{IEEEtranS}
% Generated by IEEEtranS.bst, version: 1.12 (2007/01/11)

\end{document}